\pdfoutput=1
\documentclass[11pt]{article}

\usepackage[T1]{fontenc}
\usepackage[utf8]{inputenc}
\usepackage{lmodern}
\usepackage[margin=1in]{geometry}
\usepackage{amsmath,amssymb}
\usepackage{graphicx}
\usepackage{booktabs}
\usepackage{longtable}
\usepackage{array}
\usepackage{calc}
\usepackage{ragged2e}
\usepackage[font=small,labelfont=bf]{caption}
\usepackage{float}
\usepackage{xcolor}
\usepackage[numbers,sort&compress]{natbib}
\usepackage[hidelinks]{hyperref}
\usepackage{microtype}
\usepackage{placeins}
\usepackage{needspace}

\begin{document}

\begin{center}
    {\LARGE\bfseries
    TyxonQ: A Workflow Layer for Quantum Computing in Chemistry
    \par}
 
    \vspace{6pt}
 
    {\large
    Coverage, Industrial Access, and Open Integration
    \par}
 
    \vspace{14pt}

    Xiaoran Li$^{1,2,3,*}$, 
    Zhi Yin$^{1,2,*}$,
    Zhupeng Han$^{1,2}$,
    Zhengquan Pan$^{1,2,4}$,
    Kangyu Zheng$^{5}$,

    
    \textit{$^1$ QureGenAI(Suzhou) Biotechnology Co., Ltd., Suzhou, Jiangsu, China}

    \textit{$^2$ Shanghai Qubechip Quantum technology Co., Ltd., Shanghai, Shanghai, China}
    
    \textit{$^3$ QureGenAI Joint Lab, China Pharmaceutical University, Nanjing, Jiangsu, China}
    
    \textit{$^4$ School of Physics, Xi’an Jiaotong University, Xi’an, China}

    \textit{$^5$ Department of Computer Science and Engineering, The Chinese University of Hong Kong, Hong Kong SAR, China}
    



    \vspace{10pt}

    {\small
    
    $^*$ Corresponding authors. E-mail: lixiaoran@quregenai.com; yinzhi@quregenai.com;
    }
\end{center}


\begin{abstract}
Quantum computational chemistry is often framed around a hybrid quantum solver.
In practical applications, however, bottlenecks to adoption also lie in the
surrounding multiscale workflow. A molecular problem may combine classical
molecular mechanics and dynamics, conventional electronic-structure calculations,
and a quantum solver for selected orbitals or embedded fragments. This division
of work is particularly relevant when a chemically important electronic subspace
is strongly correlated. The size of the electronic Hilbert space and the qubit,
noise, and circuit-depth constraints of noisy intermediate-scale quantum (NISQ)
devices make active-space and embedding choices central to a feasible quantum
calculation. The resulting energies and gradients must remain consistent with
the surrounding classical simulation.

We present TyxonQ, a workflow layer that connects these levels of treatment rather
than addressing only the hybrid quantum algorithm. Four integration interfaces
link PySCF-compatible energy/force scanners and embedding, solver adapters,
simulator and hardware backends, and classical consumers. The implementation
combines QM/MM and fragment embedding with conventional CCSD/FCI fragment solvers,
variational and sampling-based quantum methods, and adapters for ASE, i-PI,
OpenMM, LAMMPS (through i-PI), and MDI. We describe the contracts and audit coverage,
and evaluate correctness, switching cost, reproducibility, and artifact
portability on small chemistry workflows, distinguishing executed paths from
source-level support. The purpose of TyxonQ is to support incremental
integration of quantum calculations into existing computational chemistry
workflows, including those used in industry, while retaining appropriate
classical treatments elsewhere.
\end{abstract}

\section{Introduction}\label{introduction}
Quantum computing for chemistry \citep{cao2019,mcardle2020} is often discussed in terms of Hamiltonians, ans\"atze, and energy estimation. Practical molecular applications also require a multiscale division of work. Molecular mechanics (MM) can describe an environment, classical molecular dynamics (MD) can propagate nuclei, and conventional electronic-structure methods can treat suitable QM regions or fragments. A selected orbital subspace or embedded fragment may then be assigned to a quantum solver when its classical treatment becomes limiting. These are coupled modeling choices, not a requirement that every calculation use all three levels or that an entire molecule run on a quantum device.

Strong correlation in a chemically important electronic subspace is one motivation for considering this division, but it does not by itself establish a need for quantum computation. Classical methods remain appropriate where they provide sufficient accuracy at feasible cost; for example, conventional coupled-cluster singles and doubles (CCSD) can treat suitable predominantly single-reference regions. Where quantum treatment is explored, the electronic Hilbert space and the qubit, noise, and circuit-depth limits of noisy intermediate-scale quantum (NISQ) devices constrain the active space. QM/MM and wavefunction embedding address complementary choices of environment and electronic subsystem; they are not interchangeable approximations. Choosing these spaces and assessing the resulting model remain chemical tasks, rather than decisions made automatically by the framework.

TyxonQ organizes the interfaces within this multiscale workflow around four hand-offs. Geometry, charge, spin, and environment data define a QM region and its active-space or fragment problem (H1); that problem reaches an appropriate classical or quantum solver adapter (H2); the quantum branch is evaluated numerically or through a simulator or hardware backend (H3); and the assembled energy and nuclear gradient reach a classical optimizer or MD engine (H4). The classical portions remain outside the quantum-execution branch. Units, orbital and bit ordering, gradient availability, and process topology must remain consistent across these boundaries. A solver accurate at a single geometry is not sufficient for dynamics if its potential energy surface is non-smooth.

Existing software covers individual hand-offs well. Circuit SDKs provide construction, transpilation, simulation, and differentiable programming \citep{qiskit2024,cirq,pennylane,tensorcircuit,zhang2023tensorcircuit}; chemistry stacks provide Hamiltonians, mappings, and ans\"atze \citep{qiskit-nature,pennylane-qchem,tencirchem}; quantum--MD bridges connect solvers to ASE \citep{dononelli2026,nirmal2022}; and fragment packages combine EWF and SQD, sometimes with analytic gradients \citep{qfm,ewfgeom}. Classical engines supply force interfaces and coupling protocols \citep{ase,ipi,openmm,lammps,mdi}. The integration burden remains with the user when these components expose different conventions or a bridge is tied to one engine or solver stack.

We present TyxonQ, a workflow layer connecting classical simulation, conventional quantum chemistry, and quantum active-space or fragment calculations. Its four interfaces cover chemistry and energy/force scanners, solver integration, quantum execution, and classical consumers. PySCF supplies mean fields, integrals, and gradients; the EWF layer connects conventional CCSD/FCI fragment solvers or a custom cluster-solver callback to embedding and energy assembly. Quantum adapters supply variational and sampling-based alternatives, while ASE, i-PI, OpenMM, LAMMPS (through i-PI), and MDI retain their classical consumer roles. Qiskit, OpenFermion, and Vayesta-derived code supply established compilation, mapping, and embedding components. The scanner is the consumer-facing unit of composition; gate and pulse programs and the counts/metadata contract belong to quantum device execution. The CASCI-based SQD scanner enforces a fixed sampled subspace for force consumers.

The contribution is this multiscale integration contract, together with a coverage audit and measurements of correctness, reproducibility, switching cost, artifact portability, and deployment requirements. The evidence distinguishes executed paths from source-level support and reports the implementation boundaries alongside the capabilities.

Section~\ref{related-work} reviews prior softwares, Section~\ref{requirements} states the requirements and architectural choices, and Sections~\ref{design} and~\ref{implementation} describe the interfaces and their implementation. Section~\ref{evaluation} reports the measurements, followed by discussion in Section~\ref{discussion} and limitations in Section~\ref{limitations}.
\section{Related Work}
\label{related-work}
TyxonQ connects existing software layers rather than replacing them. Qiskit, Cirq, PennyLane, and TensorCircuit-NG provide circuit construction, compilation, simulation, or differentiable programming \citep{qiskit2024,cirq,pennylane,tensorcircuit,zhang2023tensorcircuit}. TyxonQ calls Qiskit's routing, QASM2 parsing, and object-interchange APIs. Pulse functionality has a different scope: Qiskit 2.0 removed \texttt{qiskit.pulse} \citep{qiskit-pulse-removal}, while PennyLane models differentiable Hamiltonian evolution \citep{pennylane-pulse}. TyxonQ generates pulse schedules and TQASM artifacts with OpenPulse-style declarations; generation alone does not establish external-language conformance, pulse fidelity, or calibration quality.

Qiskit Nature and PennyLane's \texttt{qml.qchem} provide molecular Hamiltonians, active spaces, mappings, and ansatz libraries \citep{qiskit-nature,pennylane-qchem}; TenCirChem focuses on efficient classical simulation of chemistry ans\"atze \citep{tencirchem,li2023tencirchem}. The interfaces reviewed here do not establish the same multi-engine energy/force contract. Qiskit Nature's \texttt{GaussianForcesDriver} supplies vibrational force constants, not an MD nuclear-force calculator. PennyLane's \texttt{optimize\_geometry} is a classical RHF/UHF utility; this restriction should not be read as applying to all PennyLane quantum geometry-optimization workflows.

The closest quantum--MD bridges connect a solver to ASE. Dononelli demonstrates a Qiskit Nature--ASE interface with central finite-difference forces and active-learning-assisted MD \citep{dononelli2026}. QM/MM embedding is discussed as an extension rather than an executed calculation. The earlier VQE-E/VQE-F interface demonstrates force-driven geometry optimization of water \citep{nirmal2022}. These works establish the force-provider role of a quantum solver; their reported workflows are centered on ASE rather than a common contract across multiple consumer engines.

Fragment packages already connect conventional quantum chemistry and quantum solvers within an embedding model. \texttt{quantum-fragment-methods} combines EWF, SQD \citep{sqd2024}, and classical solvers for protein-scale single-point calculations \citep{qfm}. \texttt{ewf-geom-opt} provides analytic gradients through a custom ASE calculator coupled to Sella; MD trajectories are left to future work \citep{ewfgeom}. Neither observation establishes a permanent absence of other functionality. TyxonQ reuses molecular RHF/UHF EWF adapted from Vayesta 1.0.1 \citep{vayesta}, with density and gradient modules adapted from MIQuLab. Its LUCJ construction follows the local unitary cluster Jastrow ansatz \citep{lucj2023} and ffsim \citep{ffsim}. These are upstream methods, not methodological contributions of this paper; the additional scope is their connection to chemistry, execution, and classical-consumer contracts.

ASE, i-PI, OpenMM, LAMMPS, and MDI supply calculators, force protocols, potentials, and coupling infrastructure on the classical side \citep{ase,ase2017,ipi,ipi2019,openmm,openmm2017,lammps,lammps2022,mdi}. Their generic interfaces still require a quantum-solver adapter to reconcile units, spin, gradients, and process topology. General workflow systems have a complementary role: AiiDA manages provenance and HPC jobs \citep{aiida2020}, FireWorks high-throughput workflows \citep{fireworks2015}, QMflows chemistry workflows \citep{qmflows2019}, and quacc ASE-based materials workflows \citep{quacc}. TyxonQ does not replace their scheduler or provenance database. It supplies the domain-specific execution contract that an orchestrator can call, with a scanner as a calculation node and compiled artifacts as provenance objects.

Table 1 summarizes these roles. The contribution is the combination of chemistry, solver, backend, and consumer interfaces under one workflow contract, rather than any individual adapter.

\Needspace{6\baselineskip}
\textbf{Table 1. Layer ownership.}

{\def\LTcaptype{none} 
\begin{longtable}[]{@{}
  >{\raggedright\arraybackslash}p{(\linewidth - 4\tabcolsep) * \real{0.3333}}
  >{\raggedright\arraybackslash}p{(\linewidth - 4\tabcolsep) * \real{0.3333}}
  >{\raggedright\arraybackslash}p{(\linewidth - 4\tabcolsep) * \real{0.3333}}@{}}
\toprule\noalign{}
\begin{minipage}[b]{\linewidth}\raggedright
Layer
\end{minipage} & \begin{minipage}[b]{\linewidth}\raggedright
Representative tools
\end{minipage} & \begin{minipage}[b]{\linewidth}\raggedright
Role in this paper
\end{minipage} \\
\midrule\noalign{}
\endhead
\bottomrule\noalign{}
\endlastfoot
Circuit SDK and transpilation & Qiskit, Cirq & reused: routing, QASM2 import, object interop \\
Differentiable programming and pulse modeling & PennyLane, TensorCircuit & not replaced; pulse is modeled at a different abstraction \\
Chemistry stacks & Qiskit Nature, PennyLane \texttt{qml.qchem}, TenCirChem & reused concepts; the contribution is the surrounding contract \\
Quantum--MD bridges & Dononelli ASE calculator, VQE-E/VQE-F & single-engine prior art; extended to multiple engines and vendors \\
Fragment-based embedding & QFM, ewf-geom-opt, Vayesta & reused (EWF); no methodological novelty claimed \\
Classical MD and QM/MM & ASE, i-PI, OpenMM, LAMMPS, MDI & consumer adapters; no replacement \\
Workflow layer & \textbf{TyxonQ} & \textbf{claimed contribution} \\
\end{longtable}
}

Table 10 in Section~\ref{discussion} compares the resulting workflow scope with the closest bridges. The comparison concerns the hand-offs covered by published interfaces, not solver accuracy or performance.
\section{Requirements and Design Space}\label{requirements}
\subsection{Workflow requirements}\label{requirements-derived-from-chemistry-and-md-workflows}

The four hand-offs impose eight requirements (Table 2). In a multiscale calculation, composability (R4) must preserve appropriate classical treatments while allowing a selected electronic subsystem to use another solver. Domain fidelity (R5) makes the QM region, units, charge, spin, active space, embedding, and gradient availability explicit. Reproducibility (R1) covers routing, bit order, and versioned provenance in the quantum branch; portability (R2) preserves the solver and consumer when execution changes; and auditability (R3) requires explicit artifacts and metadata. Cost transparency (R6) includes call overhead and dependencies, while operational security (R7) keeps credentials out of code and artifacts. Extensibility (R8) requires documented interfaces through which third-party components can join. These are design requirements rather than claims that every path is production-ready.

\textbf{Table 2. Requirements, design responses, and evidence.}

{\def\LTcaptype{none} 
\begin{longtable}[]{@{}
  >{\raggedright\arraybackslash}p{(\linewidth - 4\tabcolsep) * \real{0.3333}}
  >{\raggedright\arraybackslash}p{(\linewidth - 4\tabcolsep) * \real{0.3333}}
  >{\raggedright\arraybackslash}p{(\linewidth - 4\tabcolsep) * \real{0.3333}}@{}}
\toprule\noalign{}
\begin{minipage}[b]{\linewidth}\raggedright
Requirement
\end{minipage} & \begin{minipage}[b]{\linewidth}\raggedright
Design response
\end{minipage} & \begin{minipage}[b]{\linewidth}\raggedright
Evidence
\end{minipage} \\
\midrule\noalign{}
\endhead
\bottomrule\noalign{}
\endlastfoot
R1 Reproducibility & fixed routing seed; explicit bit order; version pinning; artifact retention & compiler tests; zero-shot configuration equality \\
R2 Portability & backend descriptor \texttt{(provider,\ device,\ shots)}; one solver call & numerical configuration check; five hardware adapters at source level \\
R3 Auditability & QASM2, QASM3, TQASM, JSON artifacts plus \texttt{result\_meta} & gate round-trip 6/6 ops; emitted TQASM (not externally validated) \\
R4 Composability & four interfaces; scanner as unit of composition & solver/method switch through one callable \\
R5 Domain fidelity & units in atomic units; charge/spin in scanner; QM/MM and periodic Ewald; frozen-subspace rule & H\(_2\)O gradient vs finite difference; frozen guard \\
R6 Cost transparency & documented dependency boundary; overhead measured against direct CASCI & Table 7 and Figure~\ref{fig:overhead} \\
R7 Operational security & per-provider token handling; no disk persistence in drivers & source inspection \\
R8 Extensibility & adapter seams and method registry; entry-point discovery planned & source; roadmap \\
\end{longtable}
}

Two of these requirements are met only in a limited sense. R7 describes the behavior of the current drivers and is not a security certification. R8 describes the interfaces; a finished plugin ecosystem does not exist yet, the current dispatch is explicit, and adding a provider still requires code changes. We return to this distinction in Section~\ref{adoption-and-roadmap}.

\subsection{Architectural choices}\label{design-space-and-rejected-alternatives}

Layered contracts were chosen over a monolithic stack so that a quantum calculation can be inserted into an existing multiscale model without replacing the surrounding classical simulation. This supports component replacement at the cost of more interfaces and documentation. A scanner-first API matches the geometry, energy, and gradient expected by chemistry consumers; a circuit API remains available for lower-level work. In the quantum device-execution branch, counts-first results retain the measurement information needed for mitigation and cross-provider normalization, rather than hiding it behind an expectation value. Conventional electronic-structure solvers do not need to pass through this measurement contract. Explicit artifacts add exporters to maintain but make quantum runs inspectable and replayable.

Reusing established integrals, gradients, routing, mappings, and embedding code keeps the contribution at the workflow level, with a dependency footprint and exposure to upstream changes (Table 3). Provider dispatch is explicit while the contracts stabilize; entry-point discovery is planned rather than implemented. Global backend and pipeline defaults simplify single-workflow use, but concurrent tasks with different defaults require separate processes. These choices distinguish the current implementation from the longer-term extension model.

The requirements are enforced where their semantics are known: geometry, units, and smoothness at the scanner; method and runtime selection at the solver; artifacts, counts, credentials, and bit order at the backend; and force conversion and process topology at the consumer. The following section describes these interfaces.
\section{Design}\label{design}
\begin{figure}[htbp]
  \centering
  \includegraphics[width=\linewidth]{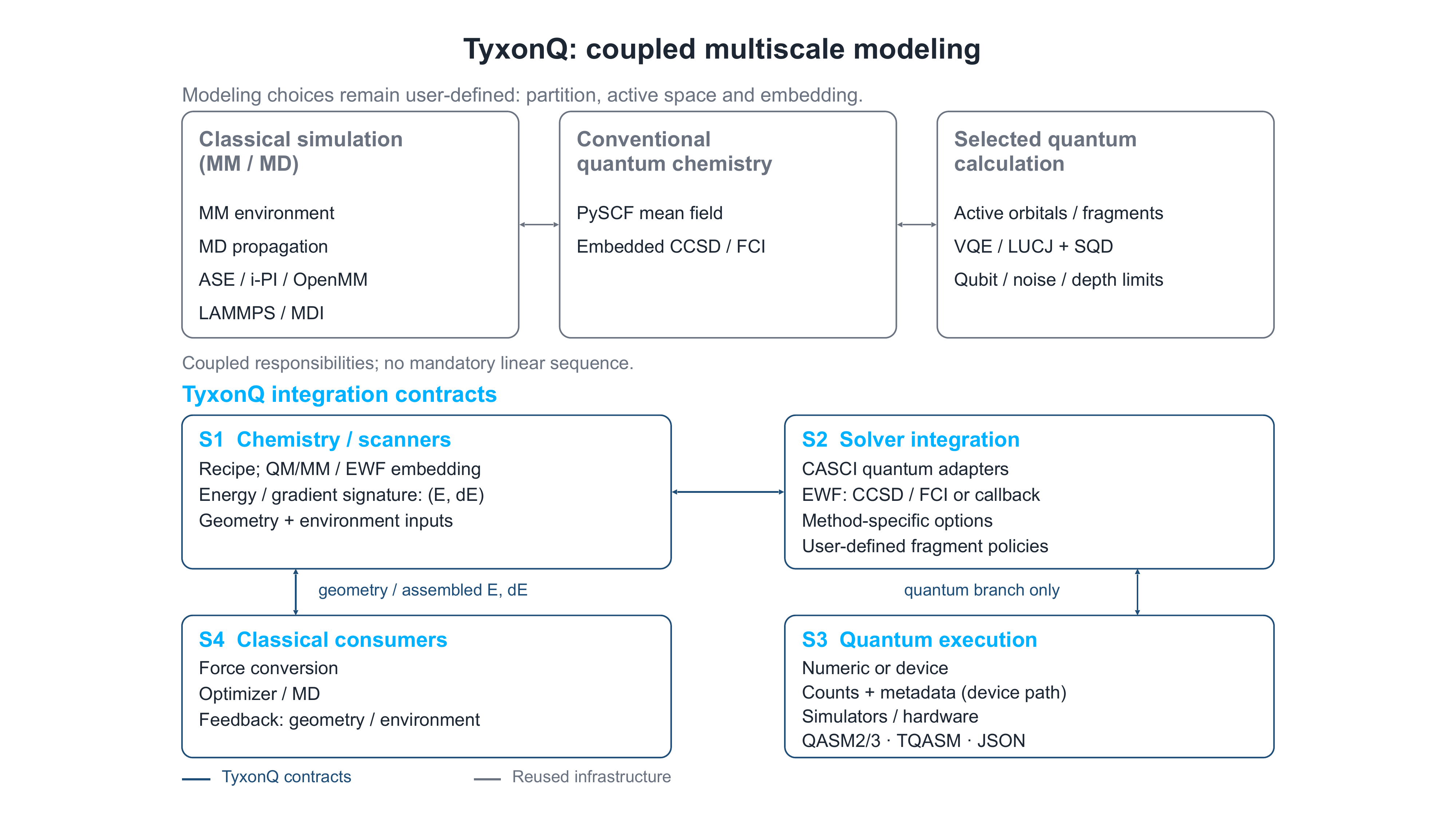}
  \caption{Multiscale treatment and the four TyxonQ integration interfaces.
  Classical MM/MD, conventional electronic structure, and selected quantum
  calculations have coupled roles rather than forming a mandatory sequence.
  The molecular partition, active space, embedding, and solver policy remain
  user-defined. S1 holds chemistry and scanner recipes; S2 connects the
  CASCI-based quantum adapters and EWF conventional or custom cluster solvers;
  S3 handles only the quantum-execution branch, with counts, metadata, and
  artifacts; S4 returns assembled energies and gradients to classical consumers.
  Blue outlines denote TyxonQ contracts and grey outlines reused infrastructure.
  Conventional CCSD does not pass through the quantum measurement contract.}
  \label{fig:surfaces}
\end{figure}

Figure~\ref{fig:surfaces} places the four interfaces within the multiscale division of work. The user specifies the molecular model, QM/MM partition where applicable, and active-space or fragment recipe. Classical mean-field and embedding calculations construct the electronic subsystem; a conventional or quantum solver supplies its contribution to energy and density assembly. Only the quantum branch uses circuit execution and normalized measurement results. A scanner returns \texttt{(E,\ dE)} to an optimizer or classical MD consumer, while the surrounding environment retains its classical treatment. The interfaces are intended to localize integration changes; they do not make different physical models equivalent.

\subsection{Chemistry and solver interfaces (S1 and S2)}\label{surface-1-chemistry-input-and-the-force-scanner}\label{surface-2-one-solver-callable-method-specific-options}

The scanner follows the PySCF nuclear-gradient signature:

\begin{verbatim}
scanner(geometry_bohr) -> (energy_hartree, gradient_hartree_per_bohr)
\end{verbatim}

It holds a recipe consisting of basis, charge, spin, active space, embedding, and method options. Construction takes a PySCF-style atom string or list; a scanner call accepts Bohr coordinates or a PySCF \texttt{Mole}. The active space is specified as \texttt{(n\_electrons,\ n\_orbitals)}, with spatial orbitals counted. The CASCI-based scanner builds an RHF reference for zero spin and ROHF otherwise; \texttt{spin} denotes \(N_\alpha-N_\beta\). Optional MM charges and periodic-Ewald data are applied before the solve. The scanner returns the QM gradient and provides a separate MM gradient for the back-reaction force. Appendix~\ref{appendix-a.-scanner-contract} specifies the arguments and units.

Solver integration preserves the consumer-facing energy/gradient signature but uses method-specific entry points. The CASCI-based \texttt{qc\_scanner} accepts quantum solver adapters, while \texttt{EWFScanner} provides a separate fragment-embedding path. EWF supplies conventional PySCF CCSD/FCI cluster solvers and accepts a \texttt{cluster\_solver} callback for another treatment, including LUCJ--SQD. The callback receives each fragment and cluster, allowing a user-defined solver policy; the framework does not infer which fragments require quantum treatment. The shared signature is therefore a compatibility boundary, not a single method selector for all classical and quantum methods.

Active-space and embedding choices specify different parts of the model. QM/MM partitions an explicit QM region from a classical environment; an active space selects electrons and orbitals within the electronic calculation; and EWF constructs fragment-plus-bath problems. They can reduce or organize the quantum workload, but the selected space must still retain the relevant chemistry. VQE-family settings use \texttt{solver\_kwargs}, SQD shots and backend settings belong to the sampler, and EWF has fragment and bath settings. The VQE runtime selects classical numerical evaluation of the ansatz or circuit submission, not conventional CCSD versus quantum UCCSD. At \texttt{shots=0}, both VQE configurations take the analytic numerical fast path; their agreement, checked in Section~\ref{eq2-contract-correctness}, is not a device-submission test.

Sampling introduces a consumer constraint. Resampling a determinant subspace at each geometry can produce a non-smooth potential energy surface; gradients evaluated within successive subspaces need not describe one fixed-subspace energy function. For \texttt{qc\_scanner(method="sqd")}, the subspace is frozen after the first geometry, and \texttt{refresh}/\texttt{adaptive} are rejected unless discontinuous behavior is explicitly accepted. This guard does not apply to \texttt{EWFScanner}; the LUCJ--SQD example rebuilds the embedding and resamples at each new geometry, so force smoothness requires separate assessment.

\subsection{Execution and artifacts (S3)}\label{surface-3-backend-interface-and-result-contract}

A backend is described by a provider, device, shot count, and provider-specific options. Drivers implement submission and result retrieval; polling and cancellation depend on the provider, and QCOS cancellation is not implemented. The common result envelope is:

\begin{verbatim}
{ "result": {bitstring: count, ...},
  "result_meta": {...},
  "uni_status": "queued" | "running" | "completed" | "failed" | ...,
  "error": null | str }
\end{verbatim}

Drivers return counts rather than expectation objects. Shared postprocessing applies measurement bases, grouping, and mitigation, keeping these semantics out of vendor drivers. Qubit 0 occupies the most significant bit at the state-vector boundary, and counts postprocessing assumes the same q0-leftmost convention (Appendix~\ref{appendix-b.-result-schema-and-bit-order}). The LUCJ--SQD example explicitly reorders spin blocks, but normalization of every hardware provider's native bit order is not established here.

Compiled programs can be retained as IR, OpenQASM 2/3, TQASM/OpenPulse, or vendor objects. \texttt{Circuit} and \texttt{PulseProgram} are peer representations with the same \texttt{device}, \texttt{run}, and \texttt{state} vocabulary. The gate-to-pulse compiler uses default decompositions or user calibrations and emits TQASM 0.2/OpenPulse when needed; gate-level execution remains the chemistry default. The compiler separates device-independent passes, dialects, and exporters so that artifacts can be inspected, archived, and replayed. This concerns program representation, not pulse fidelity.

\subsection{Classical consumers (S4)}\label{surface-4-consumer-adapters}

Five adapters consume the same energy/gradient contract: an ASE calculator, an i-PI socket driver, an OpenMM embedding potential, a LAMMPS path through i-PI's \texttt{fix\ ipi}, and an MDI engine. Each converts units and the gradient sign and handles its own process topology. LAMMPS uses three processes (LAMMPS, i-PI, and TyxonQ), not a native TyxonQ \texttt{fix\ qmmm}. Moving embedding environments use \texttt{set\_mm\_charges} and the scanner's separate MM gradient.

These consumers, MD propagation, and the MM force field remain classical. Conventional electronic-structure steps also remain classical computations. Quantum resources, when used, enter only through the selected active-space or fragment solver; the consumer receives the assembled energy and gradient rather than a quantum circuit result. This is the role of the adapter contract in a multiscale workflow: to permit a change in electronic treatment without replacing the MD engine, subject to the gradient and smoothness requirements of the chosen method.

\FloatBarrier
\section{Implementation}\label{implementation}
\subsection{Modules and dependencies}\label{repository-and-module-boundaries}\label{dependency-boundary}

The codebase is a single Python package with a thin core and a large domain layer. At the time of writing it contains 242 Python files (51,636 newline-counted lines) in \texttt{src/tyxonq}, plus 133 Python example files and 158 test files; these counts do not imply that every example was executed. The lines are distributed across modules as follows:

{\def\LTcaptype{none} 
\begin{longtable}[]{@{}
  >{\raggedright\arraybackslash}p{(\linewidth - 4\tabcolsep) * \real{0.3000}}
  >{\raggedleft\arraybackslash}p{(\linewidth - 4\tabcolsep) * \real{0.4000}}
  >{\raggedright\arraybackslash}p{(\linewidth - 4\tabcolsep) * \real{0.3000}}@{}}
\toprule\noalign{}
\begin{minipage}[b]{\linewidth}\raggedright
Module
\end{minipage} & \begin{minipage}[b]{\linewidth}\raggedleft
Lines
\end{minipage} & \begin{minipage}[b]{\linewidth}\raggedright
Responsibility
\end{minipage} \\
\midrule\noalign{}
\endhead
\bottomrule\noalign{}
\endlastfoot
\texttt{applications/} & 24,455 & chemistry, spin models, QML adapters \\
\texttt{compiler/} & 7,792 & gate passes, platform dialects, pulse compiler, exporters \\
\texttt{libs/} & 7,570 & gate kernels, circuit templates, Hamiltonian encoding, optimizers \\
\texttt{devices/} & 4,569 & simulators, hardware drivers, topology, noise channels \\
\texttt{core/} & 2,799 & \texttt{Circuit} and \texttt{PulseProgram} IR \\
\texttt{numerics/} & 2,157 & \texttt{ArrayBackend} protocol and NumPy, PyTorch, CuPyNumeric implementations \\
\texttt{postprocessing/} & 1,097 & counts\(\rightarrow\)expectation, readout mitigation, error mitigation \\
\texttt{cloud/}, \texttt{plugins/}, \texttt{visualization/} & 313 & cloud facade, plugin placeholder, drawing \\
Package-root modules & 884 & configuration, waveforms, utilities, package initialization \\
\end{longtable}
}

Modules outside \texttt{applications/} account for 27,181 lines, including numerical kernels in \texttt{libs/}. The domain layer imports the framework, not the reverse at module import time. Circuit construction therefore does not load the chemistry application layer. Table 3 separates native workflow code from reused numerical components, including PySCF integrals and gradients \citep{pyscf2020} and OpenFermion mappings \citep{openfermion2020}. This import boundary should not be confused with minimal installation requirements.

\Needspace{8\baselineskip}
\textbf{Table 3. Dependency boundary.}

{\def\LTcaptype{none} 
\begin{longtable}[]{@{}
  >{\raggedright\arraybackslash}p{(\linewidth - 4\tabcolsep) * \real{0.3333}}
  >{\raggedright\arraybackslash}p{(\linewidth - 4\tabcolsep) * \real{0.3333}}
  >{\raggedright\arraybackslash}p{(\linewidth - 4\tabcolsep) * \real{0.3333}}@{}}
\toprule\noalign{}
\begin{minipage}[b]{\linewidth}\raggedright
Dependency
\end{minipage} & \begin{minipage}[b]{\linewidth}\raggedright
Used for
\end{minipage} & \begin{minipage}[b]{\linewidth}\raggedright
Not used for
\end{minipage} \\
\midrule\noalign{}
\endhead
\bottomrule\noalign{}
\endlastfoot
Qiskit & SABRE routing (\texttt{transpile}), QASM2 import, \texttt{QuantumCircuit} interop, QCOS object path & IR definition, simulation, chemistry \\
PySCF & \texttt{Mole}, SCF, integrals, gradients, embedded CCSD/FCI, QM/MM and periodic Ewald & circuits, execution, result normalization \\
OpenFermion & fermionic operators, JW/BK/parity mappings, Hamiltonian interchange & simulation \\
ffsim (derived) & LUCJ circuit construction (Givens/DF) & SQD diagonalization \\
Vayesta 1.0.1 (derived) & EWF fragmentation, bath construction, cluster energy & scanner, MD adapters \\
MIQuLab (derived) & EWF density assembly and frozen-bath gradient & EWF fragmentation \\
renormalizer & spin-boson and pyrazine dynamics & chemistry workflows \\
PyTorch / NumPy & array backend, autograd & task management \\
vendor SDKs & cloud submission & local simulation \\
\end{longtable}
}

\subsection{Execution and chemistry integration}\label{execution-pipeline}\label{chemistry-stack}

A \texttt{Circuit} stores operations as tuples. Compilation follows typed steps:

\begin{verbatim}
AutoMeasure -> GatesTransform -> decomposition
  -> Routing -> MergePrune -> Lightcone
\end{verbatim}

These steps insert measurements, translate the basis, decompose gates, route with a fixed Qiskit seed, and prune local or light-cone operations. Platform dialects delegate to the common core; QASM2/3 exporters are native, while QASM2 import uses Qiskit. Execution resolves a driver from \texttt{(provider,\ device)}, submits a task, normalizes its payload, and applies shared postprocessing. Functional and chain APIs use the same compiler and retain compile metadata or the artifact. The mitigation modules provide readout correction, Pauli aggregation, zero-noise extrapolation, dynamical decoupling, randomized compiling, and classical shadows. Dynamical decoupling and randomized compiling include gate-sequence transformations before execution rather than only counts postprocessing. Of 785 collected core tests, 763 passed, one failed because the optional \texttt{cqlib} package was unavailable, and 21 were skipped. This result is not a fully passing conformance suite.

\begin{figure}[htbp]
  \centering
  \includegraphics[width=\linewidth]{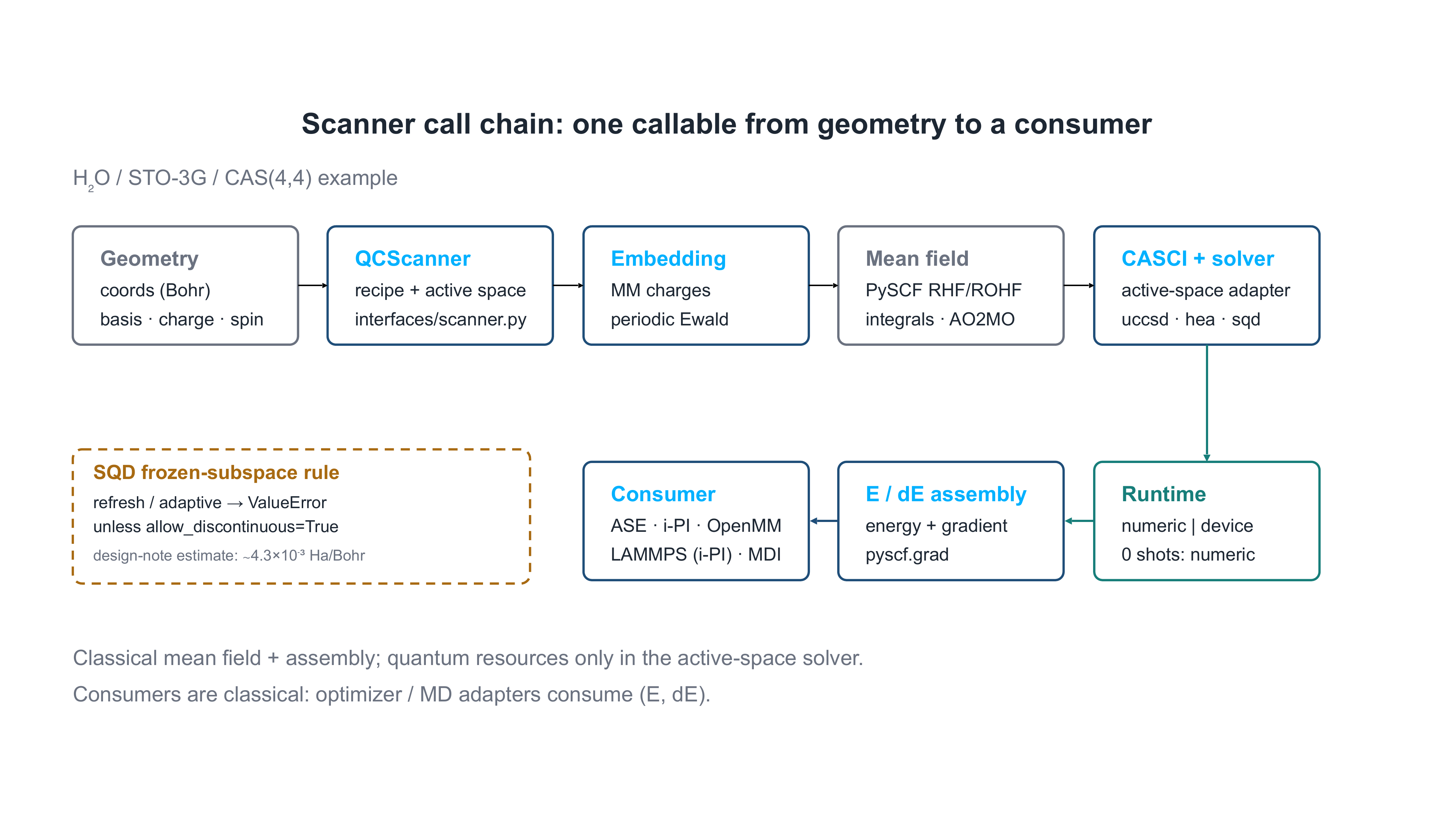}
  \caption{Scanner call chain. A geometry in Bohr and a chemistry recipe enter
  \texttt{QCScanner}; optional MM charges and a periodic lattice are applied
  before the classical PySCF mean field. A CASCI object delegates the active-space
  solve through its solver slot to a UCCSD, HEA, or SQD adapter. Mean-field,
  energy and nuclear-gradient assembly, and the consumer remain classical;
  quantum resources, when used, enter only at the active-space solve. This figure
  shows the CASCI-based path, not the separate EWF CCSD/FCI entry point. The dashed
  amber box states the frozen-subspace rule for the CASCI-based SQD scanner:
  \texttt{refresh}/\texttt{adaptive} modes are rejected unless the caller accepts
  a discontinuous surface.}
  \label{fig:scanner}
\end{figure}

The chemistry layer groups VQE-family, LUCJ, SQD, SKQD, and EWF methods in \texttt{algorithms/}, and scanner and consumer adapters in \texttt{interfaces/}. The CASCI-based scanner installs an \texttt{as\_pyscf\_solver} adapter to reuse PySCF's gradient machinery. Its method registry contains \texttt{sqd}, \texttt{uccsd}, \texttt{rouccsd}, and \texttt{hea}, with \texttt{sqd} as the default. Separately, EWF's built-in cluster dispatch calls conventional PySCF CCSD or FCI and returns energies and reduced density matrices; a \texttt{cluster\_solver} callback can provide another fragment treatment. \texttt{EWFScanner} connects the density-assembly path to a geometry-dependent energy and frozen-bath approximate gradient. Thus classical CCSD compatibility is available through EWF, not through a \texttt{method="ccsd"} option in \texttt{qc\_scanner}. Figure~\ref{fig:scanner} shows the CASCI-based call chain. A minimal numerical UCCSD call is:

\begin{minipage}{\linewidth}
\begin{verbatim}
from pyscf import gto
from tyxonq.applications.chem.interfaces import qc_scanner

water = "O 0 0 0.1173; H 0 0.7572 -0.4692; H 0 -0.7572 -0.4692"
mol = gto.M(atom=water, basis="sto-3g", unit="Angstrom", verbose=0)
coords_bohr = mol.atom_coords()
scan = qc_scanner(
    water, basis="sto-3g", active_space=(4, 4), method="uccsd",
)
E, dE = scan(coords_bohr)
\end{verbatim}
\end{minipage}

Here \texttt{active\_space=(4, 4)} means four electrons in four spatial orbitals, and the call returns Hartree and Hartree/Bohr. A VQE device call adds \texttt{solver\_kwargs} with \texttt{runtime="device"}, provider, device, and shots. For H\(_2\)/CAS(2,2), the evaluation uses the \texttt{simulator} provider and \texttt{statevector} device at \texttt{shots=0}. An ASE user assigns \texttt{TyxonQCalculator} with the same recipe to \texttt{atoms.calc}; the adapter converts the energy to eV and the negative gradient to eV/\AA{}. The measured contract checks are collected in Section~\ref{eq2-contract-correctness}, rather than repeated as tutorial steps.

\subsection{Packaging and extension}\label{packaging-and-operations}\label{extension-surface-current-state-and-roadmap}

The package installs from PyPI or source for Python 3.10--3.12. Optional extras cover MD, geometry optimization, spin models, vendor SDKs, and SBD; the base installation includes Qiskit, PySCF, OpenFermion, renormalizer, and PyTorch. Importing the package loads PyTorch even for NumPy execution (about 233 MB peak RSS in the task-level measurement). Credentials are supplied through the API or environment and are not persisted by the drivers. The cloud facade exposes submission and device listing. A separate classical client provides request wrappers for conventional electronic-structure calculations, including CCSD; it sends tasks to a classical service rather than a quantum backend. This client interface is source-level support here, not a verified remote CCSD service or a general CCSD force-scanner path. Sphinx documentation and executable examples accompany the tests.

Extensions currently use adapter modules and explicit dispatch. A solver registers a method name and exposes the numerical adapter; a backend implements submission and a raw result payload, with a compiler dialect if needed; a consumer converts the shared energy/gradient contract. Adding a provider still requires editing the resolver, and a new dialect requires editing compiler dispatch. Entry-point discovery, packaged conformance tests, and versioned contracts are planned (Section~\ref{adoption-and-roadmap}). The open-integration claim describes the separation of interfaces, not a finished plugin ecosystem.

\FloatBarrier
\section{Evaluation}\label{evaluation}
We assess five workflow properties: coverage (EQ1), scanner correctness (EQ2), reproducibility and switching (EQ3), artifact portability (EQ4), and deployment coverage (EQ5). The tests isolate interfaces needed to combine classical chemistry preparation, embedded electronic solves, and classical consumers. They do not demonstrate a strongly correlated application or establish when a quantum solver should replace a classical one. Energies use Hartree (Ha), gradients Ha/Bohr, and timings the environment pinned in Appendix~\ref{appendix-f.-reproduction}. The measurements concern interface behavior and integration cost, not solver rankings or quantum advantage. Table 4 collects the basic reference-workflow checks; the following subsections give the protocols and additional cases. ``Call-site change'' refers to the workflow author's code, excluding adapter internals.

\Needspace{17\baselineskip}
\textbf{Table 4. One workflow with four boundaries.}

{\def\LTcaptype{none} 
\begin{longtable}[]{@{}
  >{\raggedright\arraybackslash}p{(\linewidth - 6\tabcolsep) * \real{0.2500}}
  >{\raggedright\arraybackslash}p{(\linewidth - 6\tabcolsep) * \real{0.2500}}
  >{\raggedright\arraybackslash}p{(\linewidth - 6\tabcolsep) * \real{0.2500}}
  >{\raggedright\arraybackslash}p{(\linewidth - 6\tabcolsep) * \real{0.2500}}@{}}
\toprule\noalign{}
\begin{minipage}[b]{\linewidth}\raggedright
Step
\end{minipage} & \begin{minipage}[b]{\linewidth}\raggedright
Call-site change
\end{minipage} & \begin{minipage}[b]{\linewidth}\raggedright
Result
\end{minipage} & \begin{minipage}[b]{\linewidth}\raggedright
Evidence
\end{minipage} \\
\midrule\noalign{}
\endhead
\bottomrule\noalign{}
\endlastfoot
Numerical UCCSD scanner & none & \texttt{E\ =\ -74.9704540905} Ha; O \(z\) gradient \(-0.0740153863\) Ha/Bohr & H\(_2\)O/CAS(4,4) \\
Device-configured scanner & one options dict & H\(_2\) difference \texttt{0.0}; both calls numerical at zero shots & H\(_2\)/CAS(2,2) \\
ASE consumer & one adapter assignment & \texttt{E\ =\ -74.9704547011} Ha; \(F_{\max}=0.07401539\) Ha/Bohr (atomic norm) & H\(_2\)O/CAS(4,4) \\
Artifacts & one compile output selector & QASM2, QASM3, JSON round-trip 6/6 ops & 3-qubit circuit \\
\end{longtable}
}

\subsection{Coverage and deployment (EQ1 and EQ5)}\label{eq1-workflow-coverage}\label{eq5-deployment-surface}

Table 5 lists eight implemented stages and separates executed tests from source, unit-test, and author-archive evidence. An implemented path is not necessarily validated at production scale. Larger QM/MM and periodic-Ewald cases and the optional engines other than ASE were not reproduced in this environment; these evidence boundaries are discussed in Section~\ref{limitations}.

\textbf{Table 5. Workflow coverage.}

{\def\LTcaptype{none} 
\begin{longtable}[]{@{}
  >{\raggedright\arraybackslash}p{(\linewidth - 6\tabcolsep) * \real{0.2500}}
  >{\raggedright\arraybackslash}p{(\linewidth - 6\tabcolsep) * \real{0.2500}}
  >{\raggedright\arraybackslash}p{(\linewidth - 6\tabcolsep) * \real{0.2500}}
  >{\raggedright\arraybackslash}p{(\linewidth - 6\tabcolsep) * \real{0.2500}}@{}}
\toprule\noalign{}
\begin{minipage}[b]{\linewidth}\raggedright
Stage
\end{minipage} & \begin{minipage}[b]{\linewidth}\raggedright
Implementation
\end{minipage} & \begin{minipage}[b]{\linewidth}\raggedright
External boundary
\end{minipage} & \begin{minipage}[b]{\linewidth}\raggedright
Evidence in this study
\end{minipage} \\
\midrule\noalign{}
\endhead
\bottomrule\noalign{}
\endlastfoot
Classical chemistry & PySCF \texttt{Mole}, mean field; EWF CCSD/FCI solvers & PySCF & mean field and FCI reference executed; CCSD source \\
Embedding / QM-MM & EWF (Vayesta-adapted), electrostatic and periodic Ewald & PySCF \texttt{qmmm} & electrostatic and EWF cases executed; periodic archive \\
Solvers & UCC/UCCSD/HEA family; LUCJ+SQD/SKQD; EWF & OpenFermion, ffsim-derived LUCJ & UCCSD, HEA, EWF--LUCJ--SQD cases executed; others source/tests \\
Execution & statevector, density matrix, MPS; five hardware providers & vendor SDKs & simulators executed; hardware source \\
Mitigation / postprocessing & counts\(\rightarrow\)energy, readout mitigation, ZNE, classical shadows; pre-execution DD/RC & none & source + unit tests \\
Downstream & ASE, i-PI, OpenMM, LAMMPS via i-PI, MDI & classical engines & ASE executed; other engines source/author archive \\
Artifacts & QASM2, QASM3, TQASM, JSON IR & Qiskit (QASM2 import) & executed \\
Packaging & pip, optional extras, Python 3.10--3.12 & dependencies & existing Python 3.12 environment executed; fresh install not tested \\
\end{longtable}
}

Table 6 lists three credential-free simulators and five hardware submission/result adapters. Availability is source-level evidence, not hardware validation or a claim that every lifecycle operation is implemented. The IBM skeleton is excluded from the count. Packaging and classical offloading are described in Section~\ref{packaging-and-operations}; the scope and installation limits are collected in Section~\ref{limitations}.

\textbf{Table 6. Execution and deployment coverage.}

{\def\LTcaptype{none} 
\begin{longtable}[]{@{}
  >{\raggedright\arraybackslash}p{(\linewidth - 4\tabcolsep) * \real{0.3333}}
  >{\raggedright\arraybackslash}p{(\linewidth - 4\tabcolsep) * \real{0.3333}}
  >{\raggedright\arraybackslash}p{(\linewidth - 4\tabcolsep) * \real{0.3333}}@{}}
\toprule\noalign{}
\begin{minipage}[b]{\linewidth}\raggedright
Execution target
\end{minipage} & \begin{minipage}[b]{\linewidth}\raggedright
Status
\end{minipage} & \begin{minipage}[b]{\linewidth}\raggedright
Notes
\end{minipage} \\
\midrule\noalign{}
\endhead
\bottomrule\noalign{}
\endlastfoot
statevector / density matrix / MPS & available & density matrix supports Kraus channels \\
tyxonq (homebrew\_s2) & submit/result adapter & token required \\
QCOS (China Mobile ecloud) & submit/result adapter & credentials required \\
Quafu & submit/result adapter & token required \\
LQCloud & submit/result adapter & SDK/token required \\
GuoDun & submit/result adapter & QCIS; SDK required \\
IBM & skeleton & excluded from the count \\
Classical cloud & client available & HF/post-HF offload \\
\end{longtable}
}

\subsection{Scanner correctness and reproducibility (EQ2 and EQ3)}\label{eq2-contract-correctness}\label{eq3-reproducibility-and-switching}\label{consumer-check-ase-end-to-end}

We exercise the scanner on two systems: H\(_2\)O/STO-3G/CAS(4,4) with UCCSD, and H\(_2\)/STO-3G/CAS(2,2) for the numeric--device comparison.

For H\(_2\)O, the scanner returns \(E=-74.9704540905\) Ha. A CASCI calculation in the same active space gives \(-74.9704543778\) Ha, a difference of \(+2.9\times10^{-7}\) Ha. The analytic O \(z\) gradient is \(-0.0740153863\) Ha/Bohr; a central finite difference with a \(10^{-3}\) Bohr displacement gives \(-0.0740155408\) Ha/Bohr, differing by \(1.54\times10^{-7}\) Ha/Bohr. The tutorial also returns finite energies and gradients at four successive O--H displacements. The observed ordering \(E_{\mathrm{HEA}} \ge E_{\mathrm{UCCSD}} \ge E_{\mathrm{CASCI}}\) is specific to this example, not a general ordering of the two ansatz families. These results are consistency checks, not an isolated measurement of adapter bias.

For H\(_2\)/STO-3G/CAS(2,2), the numerical configuration and the device-configured call (\texttt{simulator}, \texttt{statevector}, \texttt{shots=0}) both give \(-1.1372838344885008\) Ha and identical gradients (maximum difference \(0.0\)). At zero shots, the UCCSD implementation uses the analytic numerical fast path in both configurations. The equality therefore tests configuration compatibility, not device submission, compilation, or count normalization. Routing determinism is checked separately in the core compiler tests.

\begin{figure}[htbp]
  \centering
  \includegraphics[width=0.86\linewidth]{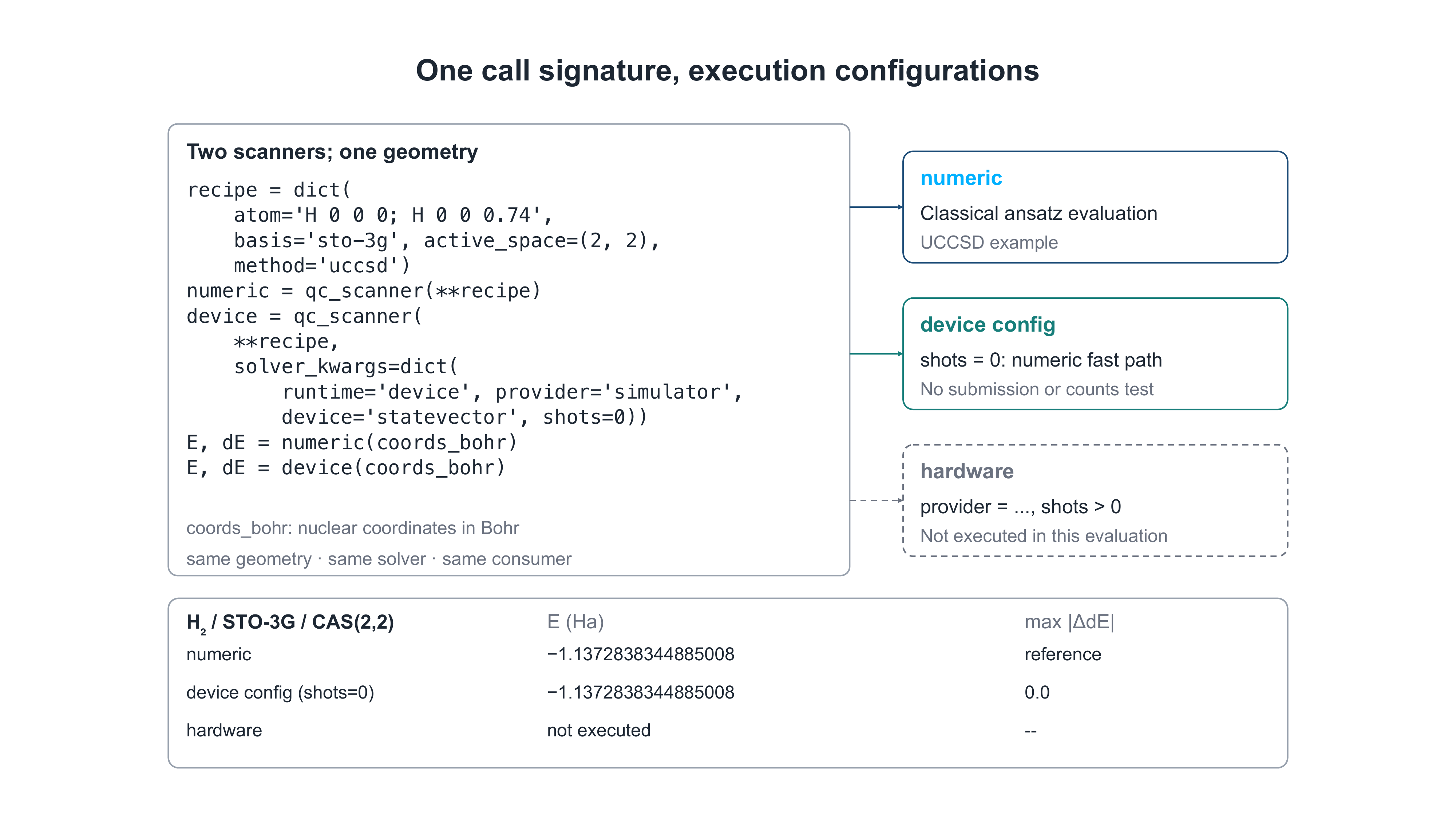}
  \caption{One call site and its execution configurations. The change is the
  \texttt{solver\_kwargs} dictionary; geometry, solver call, and consumer remain
  identical. The H$_2$/STO-3G/CAS(2,2) numeric and device-configured
  (\texttt{shots=0}) calls return the same energy and gradient because both use
  the numerical fast path. The dashed hardware branch was not executed;
  positive-shot device submission is not validated by this equality.}
  \label{fig:runtime}
\end{figure}

The device-configured call adds four option keys (runtime, provider, device, and shots) without changing the geometry or consumer (Figure~\ref{fig:runtime}). The numerical and device-configured calls take 6.51 and 6.57 ms, respectively (medians of five calls); these are two timings of the numerical fast path. Provider switching was not tested end to end on hardware here. Bit-order conventions must also be explicit: Qiskit's computational-basis index places qubit 0 at the least significant bit, whereas Cirq uses the supplied \texttt{qubit\_order} from most to least significant \citep{cirq}. TyxonQ uses qubit 0 at the most significant bit.

The ASE adapter was checked on H\(_2\)O/STO-3G/CAS(4,4). Its energy converted back from eV is \(-74.97045470108468\) Ha, differing from the direct scanner result by \(6.11\times10^{-7}\) Ha; the residual is consistent with the adapter's conversion constants. The maximum atomic force norm is 0.07401539 Ha/Bohr, and forces have the opposite sign to the nuclear gradient. Five BFGS steps reduce this norm from 3.8060 to 0.01378 eV/\AA{}, meeting the 0.05 eV/\AA{} criterion. The median measured step takes 20.66 ms. This tests unit conversion and force consumption in optimization, not an MD trajectory or a chemical prediction.

\subsection{Artifact portability (EQ4)}\label{eq4-artifact-portability}

\begin{figure}[htbp]
  \centering
  \includegraphics[width=\linewidth]{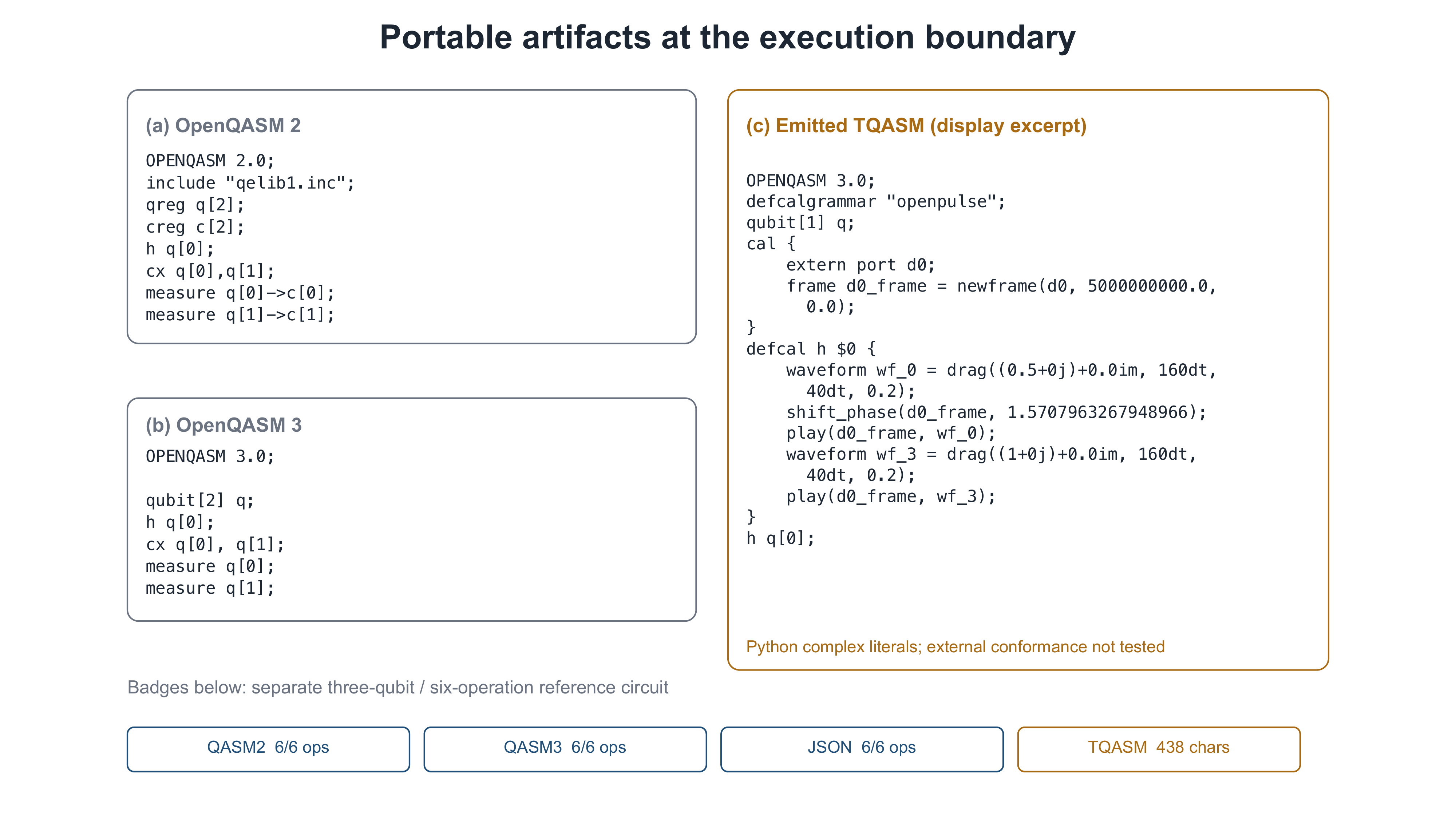}
  \caption{Portable artifacts. (a, b) A two-qubit Bell circuit exported to
  OpenQASM 2 and OpenQASM 3. (c) An excerpt of the emitted TQASM artifact for
  a one-qubit H gate (438 characters in full). It retains the compiler's Python-style complex literals;
  external OpenPulse conformance was not established.
  The badges report the round-trip results
  for the three-qubit circuit in Section~\ref{eq4-artifact-portability}:
  QASM2, QASM3, and JSON IR each recover the original six operations.}
  \label{fig:artifacts}
\end{figure}

The three-qubit reference circuit (an \(H\), a CNOT, an \(R_z\), and three measurements) round-trips through QASM2, QASM3, and JSON IR with all six operations recovered. Appendix~\ref{appendix-e.-artifact-examples} retains the export/import example. We also sweep seven circuits from 4 to 16 qubits (two layers of RX/RZ plus a CNOT chain, with measurements) and export each to QASM2, QASM3, and JSON. All 21 round trips reproduce the original operation list exactly. Artifact sizes range from 674 to 2,684 bytes for QASM2, 626 to 2,581 for QASM3, and 702 to 2,777 for JSON; the first QASM2 export takes 7.92 ms including cold-start overhead, while the remaining QASM2 exports and all QASM3 exports take at most 0.13 ms in this run. A one-qubit \(H\) circuit compiled with \texttt{use\_pulse} produces a 438-character TQASM artifact with \texttt{defcalgrammar\ "openpulse"}, a frame declaration, and a DRAG calibration (Figure~\ref{fig:artifacts}). The emitted amplitude includes the Python-style complex literal \texttt{(0.5+0j)}, so this output is retained as a compiler artifact, not validated OpenPulse source for external replay. Pulse-language conformance and device acceptance require separate tests.

The dependency boundary is explicit: QASM2 import uses Qiskit, while QASM3 export/import and the QASM2 exporter are native. The artifact layer is therefore not free of external dependencies.

\subsection{Workflow overhead}\label{workflow-overhead-relative-to-a-direct-classical-reference}

Table 7 compares scanner UCCSD with direct PySCF CASCI for energy plus analytic nuclear gradient in the same active space. The protocol uses H\(_2\), LiH, and H\(_2\)O in STO-3G, one warm-up, and the median of three calls on the pinned environment. LiH/CAS(2,3) is excluded because its reference gradient is numerically singular at this geometry; LiH/CAS(2,2) is used instead. The comparison measures the combined scanner and variational-solver cost, not like-for-like solver efficiency.

\Needspace{9\baselineskip}
\textbf{Table 7. Scanner path vs direct CASCI (energy + nuclear gradient).}

{\def\LTcaptype{none} 
\begin{longtable}[]{@{}lrrrrr@{}}
\toprule\noalign{}
System & Qubits & Scanner (ms) & CASCI (ms) & Ratio & \(\Delta\)E (Ha) \\
\midrule\noalign{}
\endhead
\bottomrule\noalign{}
\endlastfoot
H\(_2\) / CAS(2,2) & 4 & 6.22 & 3.05 & 2.04\(\times\) & 4.4e-16 \\
LiH / CAS(2,2) & 4 & 9.43 & 4.07 & 2.32\(\times\) & 6.7e-11 \\
H\(_2\)O / CAS(4,4) & 8 & 19.53 & 6.18 & 3.16\(\times\) & 2.9e-7 \\
\end{longtable}
}

\begin{figure}[htbp]
  \centering
  \includegraphics[width=0.62\linewidth]{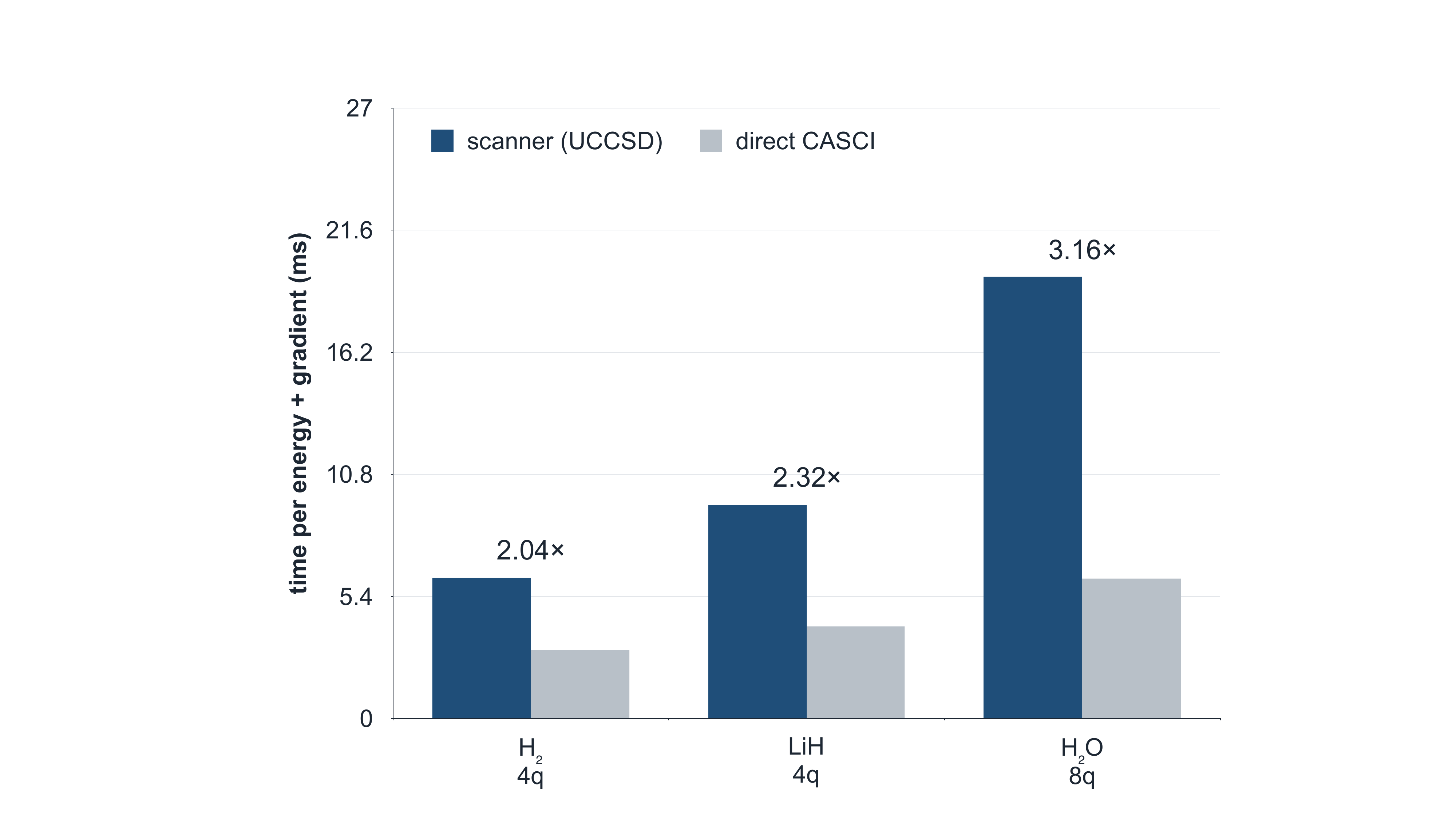}
  \caption{Workflow overhead: scanner path (UCCSD) versus direct PySCF CASCI for
  energy plus nuclear gradient on H$_2$/CAS(2,2), LiH/CAS(2,2), and
  H$_2$O/CAS(4,4). Bars are medians of three calls; labels give the ratio. The
  figure reports cost transparency, not a performance ranking.}
  \label{fig:overhead}
\end{figure}

The scanner path costs 2.04--3.16 times the direct CASCI call in this run (Figure~\ref{fig:overhead}), including mean-field updates, UCCSD parameter optimization, and gradient assembly. The direct reference reuses a prepared mean field, whereas scanner calls include its update. Because the solvers and timed work differ, these ratios do not isolate pure adapter overhead. Three small systems on one machine establish neither a scaling law nor performance leadership; the measurements make the cost of the tested workflow explicit.

\subsection{Embedding and sampling workflows}\label{case-studies}\label{workflow-case-study-qmmm-and-frozen-subspace-sampling}

Three small cases exercise electrostatic embedding, solver substitution, and fragment sampling. They are workflow checks, not production benchmarks or chemistry validation. MM denotes the molecular-mechanics point-charge environment, and mHa denotes \(10^{-3}\) Ha.

In case A, two charges of \(\pm0.4e\) at \(z=\pm3\) \AA{} are applied to water through \texttt{mm\_charges}. The UCCSD scanner energy changes from \(-74.9704540905\) to \(-74.9869058874\) Ha (\(\Delta E=-0.0164517969\) Ha), and the largest change in the QM gradient is 0.00477098 Ha/Bohr. \texttt{mm\_gradient()} returns a \(2\times3\) array with a maximum component of 0.00308252 Ha/Bohr; its negative is the MM back-reaction force. This single-process check exercises the embedded scanner, not a solvent model or a multi-engine run.

Case B changes only the method selector for the same water geometry, basis, and CAS(4,4). UCCSD gives \(-74.9704540905\) Ha and HEA \(-74.9630231371\) Ha, a difference of \(+0.00743095\) Ha. Scanner construction plus one energy/gradient call takes 36.51 ms for UCCSD and 0.359 s for HEA, timed separately for each method. Both return the same energy/gradient signature. The result tests substitution at the consumer interface; one geometry and one parameterization do not establish a general ansatz ranking, and method-specific options still differ.

Case C uses H\(_2\)/STO-3G at a bond length of 0.78 \AA{}, divides it into two hydrogen fragments, constructs baths, and solves them with a LUCJ state-vector sampler followed by SQD recovery. It uses 2048 shots, one LUCJ layer, and the \texttt{small} profile:

\begin{verbatim}
python examples/ewf_geometry.py singlepoint --molecule h2 --force-sqd
\end{verbatim}

The total energy is \(-1.1358187791\) Ha. For the first fragment, LUCJ construction takes 0.079 s, sampling 0.0004 s, SQD recovery 0.0022 s, and the fragment solve 0.092 s. Recovery converges in two iterations with \(\langle S^2\rangle=0.0\). The full process takes 2.44 s including startup and imports; the default FCI fragment solver gives the same total energy with a reported energy stage of 0.027 s. This checks the fragment/sampling/recovery chain and its metadata, not embedding scalability or agreement with a full-system solver. Table 8 collects the interface checks and costs.

\Needspace{12\baselineskip}
\textbf{Table 8. Case studies: interfaces exercised and measured results.}

{\def\LTcaptype{none} 
\begin{longtable}[]{@{}
  >{\raggedright\arraybackslash}p{(\linewidth - 8\tabcolsep) * \real{0.2000}}
  >{\raggedright\arraybackslash}p{(\linewidth - 8\tabcolsep) * \real{0.2000}}
  >{\raggedright\arraybackslash}p{(\linewidth - 8\tabcolsep) * \real{0.2000}}
  >{\raggedright\arraybackslash}p{(\linewidth - 8\tabcolsep) * \real{0.2000}}
  >{\raggedright\arraybackslash}p{(\linewidth - 8\tabcolsep) * \real{0.2000}}@{}}
\toprule\noalign{}
\begin{minipage}[b]{\linewidth}\raggedright
Case
\end{minipage} & \begin{minipage}[b]{\linewidth}\raggedright
Interfaces
\end{minipage} & \begin{minipage}[b]{\linewidth}\raggedright
Key result
\end{minipage} & \begin{minipage}[b]{\linewidth}\raggedright
Cost
\end{minipage} & \begin{minipage}[b]{\linewidth}\raggedright
Evidence
\end{minipage} \\
\midrule\noalign{}
\endhead
\bottomrule\noalign{}
\endlastfoot
A. QM/MM & scanner + embedding + MM gradient & \(\Delta\)E = $-$16.45 mHa; MM gradient max component 0.00308 Ha/Bohr & 37.48 ms including MM gradient & executed in this study \\
B. Solver substitution & solver + scanner & HEA \(-\) UCCSD = +7.43 mHa at fixed geometry & UCCSD: 36.51 ms; HEA: 0.359 s (construction + call) & executed in this study \\
C. EWF + LUCJ + SQD & fragment + sampler + selected CI & E = $-$1.1358187791 Ha; SQD converged, \(\langle S^2\rangle\) = 0.0 & fragment 0.092 s; process 2.44 s & executed in this study \\
\end{longtable}
}

The frozen-subspace guard also rejects the \texttt{refresh} and \texttt{adaptive} modes unless discontinuities are accepted, and its message cites a force-noise estimate of approximately \(4.3\times10^{-3}\) Ha/Bohr from the design note. Larger QM/MM and periodic-Ewald workflows are covered only by the TyxonQ tutorials in \texttt{examples/qmmm/} (E1--E11), not by measurements reproduced here. The executed cases establish composition on small systems; the hardware, MD-trajectory, and scale boundaries are stated in Section~\ref{limitations}.

\FloatBarrier
\section{Discussion}\label{discussion}
The evaluation supports a workflow-level result: scanner, solver, backend, and consumer conventions can be specified and checked at their boundaries. These interfaces connect parts of a multiscale calculation rather than making the quantum algorithm the whole workflow. Classical MM/MD and conventional electronic-structure methods retain their roles; a quantum solver can be introduced for a selected electronic subsystem. This separates integration failures, such as incorrect units or bit ordering, from errors in the electronic model or solver. The frozen-subspace rule illustrates how a classical consumer can constrain the admissible quantum-solver configuration. Coverage and dependency tables make the extent of this integration visible, but the small cases do not establish accuracy for strong correlation or production maturity.

For industrial users, the intended benefit is a route to introduce quantum calculations within existing computational-chemistry practice, not a requirement to replace a working classical pipeline. ``Industrial access'' refers here to common submission/result interfaces, credentials kept out of code, archivable artifacts, documented dependencies, and an explicit result convention. It does not imply service-level agreements, certification, or a lightweight installation. Portability concerns the solver and workflow at the execution boundary; actual artifact acceptance remains provider-dependent. The architectural claim would be weakened if provider changes required solver rewrites, if hardware results violated the stated semantics, or if maintaining the adapters cost more than the integration work they replace. Those questions require deployment evidence beyond this study.

\subsection{Open integration and adoption}\label{adoption-and-roadmap}

The adapter structure provides an extension point at each interface, but open integration also requires documented, versioned contracts and tests that third parties can run. Provider dispatch is currently explicit and discovery is a placeholder; these remain implementation work rather than an established plugin ecosystem. Table 9 separates implemented components from planned additions.

A conformance suite should check four kinds of behavior: scanner inputs, units, and finite-difference gradients; result fields and known bitstrings; artifact export/import; and consumer energy/force conversions. Simulator and mock-driver tests can run without credentials, although they do not replace hardware validation. Adapter templates and entry-point discovery would then reduce the need to edit the core. Versioning and deprecation rules are needed for the scanner signature, result schema, and artifact formats so that compatibility changes are explicit.

A practical adoption path starts from the user's established classical simulation and electronic-structure references. The QM region, active space, or embedded fragments are then chosen for the chemical question, and a quantum-solver candidate can be checked in simulation without replacing the surrounding model or MD engine. Hardware submission with archived artifacts is a later execution choice, not a change to the classical parts of the workflow. This route still requires validation of both the reduced chemical model and the selected solver. Further provider integrations are under evaluation and are not counted here. The goal is component-level compatibility, not a marketplace or certification program.

\Needspace{8\baselineskip}
\textbf{Table 9. Roadmap: implemented vs planned.}

{\def\LTcaptype{none} 
\begin{longtable}[]{@{}
  >{\raggedright\arraybackslash}p{(\linewidth - 4\tabcolsep) * \real{0.3333}}
  >{\raggedright\arraybackslash}p{(\linewidth - 4\tabcolsep) * \real{0.3333}}
  >{\raggedright\arraybackslash}p{(\linewidth - 4\tabcolsep) * \real{0.3333}}@{}}
\toprule\noalign{}
\begin{minipage}[b]{\linewidth}\raggedright
Item
\end{minipage} & \begin{minipage}[b]{\linewidth}\raggedright
Status
\end{minipage} & \begin{minipage}[b]{\linewidth}\raggedright
Dependency
\end{minipage} \\
\midrule\noalign{}
\endhead
\bottomrule\noalign{}
\endlastfoot
Four adapter interfaces & implemented & none \\
Five hardware providers + three simulators & implemented & vendor SDKs \\
Five consumer adapters & implemented & ASE/i-PI/OpenMM/LAMMPS/MDI \\
Scanner/method registry & implemented & PySCF \\
Result schema + bit-order contract & implemented & none \\
Conformance suite & planned & contract stabilization \\
Entry-point discovery & planned & conformance suite \\
Adapter templates & planned & conformance suite \\
Versioned contract policy & planned & maintainers' agreement \\
Public conformance badge & planned & governance \\
Additional providers (under evaluation) & planned & driver contract + conformance suite \\
\end{longtable}
}

A third-party integration should eventually be installable and testable without maintainers editing the core. The present architecture supports the separation of responsibilities; achieving that packaging and compatibility guarantee is still part of the roadmap.

\subsection{Workflow scope comparison}\label{workflow-scope-comparison}

Table 10 compares reported workflow scope with the closest prior bridges. Its last column identifies remaining work for a force-driven or QM/MM MD application, not a proof that a package lacks the corresponding functionality. The comparison combines published demonstrations and source-level adapter coverage, and involves no performance measurement.

\textbf{Table 10. Reported workflow scope and remaining integration or validation work.}

{\def\LTcaptype{none} 
\begin{longtable}[]{@{}
  >{\raggedright\arraybackslash}p{(\linewidth - 8\tabcolsep) * \real{0.2000}}
  >{\raggedright\arraybackslash}p{(\linewidth - 8\tabcolsep) * \real{0.2000}}
  >{\raggedright\arraybackslash}p{(\linewidth - 8\tabcolsep) * \real{0.2000}}
  >{\raggedright\arraybackslash}p{(\linewidth - 8\tabcolsep) * \real{0.2000}}
  >{\raggedright\arraybackslash}p{(\linewidth - 8\tabcolsep) * \real{0.2000}}@{}}
\toprule\noalign{}
\begin{minipage}[b]{\linewidth}\raggedright
Workflow
\end{minipage} & \begin{minipage}[b]{\linewidth}\raggedright
Input + solver
\end{minipage} & \begin{minipage}[b]{\linewidth}\raggedright
Execution
\end{minipage} & \begin{minipage}[b]{\linewidth}\raggedright
Consumer
\end{minipage} & \begin{minipage}[b]{\linewidth}\raggedright
Manual additions
\end{minipage} \\
\midrule\noalign{}
\endhead
\bottomrule\noalign{}
\endlastfoot
TyxonQ & PySCF; EWF/QM-MM; CCSD/FCI; UCCSD, HEA, SQD & classical solves; simulators; 5 vendor adapters & ASE; four other consumer adapters & model setup; method/engine validation \\
ASE-Qiskit VQE \citep{dononelli2026} & ASE; VQE/ADAPT & local quantum simulation & ASE optimization and MD & QM/MM; second engine \\
QFM \citep{qfm} & PySCF; EWF; SQD & IBM QPU & single-point study & force-driven workflow not evaluated \\
ewf-geom-opt \citep{ewfgeom} & PySCF; EWF; SQD/SCI & local/HPC & ASE calculator + Sella & MD-trajectory validation \\
\end{longtable}
}

The table compares integration scope, not speed or accuracy. TyxonQ's listed adapters still require configuration and method-specific validation. The separate EWF CCSD/FCI path is not a drop-in replacement for the QCScanner-based QM/MM consumer adapters; matching signatures do not establish every solver--engine combination. Published single-point or optimization demonstrations should likewise not be interpreted as blanket statements about a package's MD capabilities.
\section{Limitations}\label{limitations}
The executed quantum-solver checks use local numerical evaluation or state-vector sampling. The zero-shot VQE configuration comparison does not test device submission. TyxonQ supports hardware execution, and the authors' separate SQD application work has used TyxonQ-built LUCJ circuits submitted through a collaborator's hardware interface. Those runs will be reported separately with their device provenance and are not evidence for the provider drivers studied here. The five hardware adapters are assessed at source level; native count order, queuing, cancellation, calibration, and artifact acceptance still require provider-specific validation. The executed consumer check is ASE geometry optimization, not an MD trajectory. In prospective multiscale MD uses, nuclear propagation, the MM force field, and consumer engines remain classical; quantum resources enter only through selected electronic solves. OpenMM, i-PI, LAMMPS, MDI, larger QM/MM cases, and periodic-Ewald workflows remain source or author-archive evidence rather than independently reproduced results here. The systems (H\(_2\), LiH, H\(_2\)O) and active spaces are small; all timings come from one machine and one version set (Appendix~\ref{appendix-f.-reproduction}).

The workflow claim is scoped to molecular multiscale calculations with QM/MM or fragment embedding and selected quantum electronic solves. TyxonQ provides interfaces for a user-specified division of work, not an automatic diagnosis of strong correlation or selection of a chemically adequate active space. Molecular input is PySCF-style, and there is no general SMILES/PDB-to-3D ingestion. Periodic support is limited to QM/MM Ewald embedding; there is no plane-wave or periodic DFT, so we do not use the word ``materials'' without this qualifier. LAMMPS is reached through i-PI and not through a native \texttt{fix\ qmmm}. The embedding layer is adapted from Vayesta 1.0.1 and restricted to molecular RHF/UHF, and the density and gradient modules are adapted from MIQuLab; we treat these as reused infrastructure and claim no methodological novelty for them.

Some limitations concern the contracts and the implementation. The scanner call is shared across methods, but the option sets are method-specific, so switching methods is not zero-cost at the option level. The EWF gradient fixes the bath, projectors, and effective amplitudes; it is not the full derivative of the energy obtained by rebuilding and re-solving the embedding at a displaced geometry. Its density-gradient scanner is restricted to closed-shell nonperiodic systems and does not support PCM or QM/MM wrappers. PCM--SQD is a fixed-orbital energy/density workflow, not a nuclear-gradient scanner. VQE convergence and a frozen sampled space do not alone establish force smoothness or MD stability. TQASM generation is tested, but its complex-number serialization needs correction before external OpenPulse conformance can be claimed. The author archive documents a periodic-Ewald MM back-reaction bias of approximately \(4.3\times10^{-5}\) Ha/Bohr. The IR has no formal version or migration mechanism, and backend and pipeline defaults are process-global. Provider dispatch is explicit and plugin discovery is a placeholder, so adding a backend currently requires editing the resolver. The base install requires Qiskit, PySCF, OpenFermion, renormalizer, and PyTorch, and importing the package loads PyTorch even when NumPy is selected (approximately 233 MB peak RSS). Finally, the paper does not benchmark solver accuracy, speed, or memory against other frameworks.

Four additions would strengthen the evidence: (i) hardware runs through at least two of the provider drivers, with the result-format and bit-order checks of Appendix~\ref{appendix-b.-result-schema-and-bit-order}; (ii) an i-PI workflow and an ASE MD trajectory executed in the evaluation environment; (iii) a conformance suite that third parties can run against a new adapter; and (iv) independent replication of the scanner contract on another machine and version set. Until these are available, the claims of this paper remain architectural and contract-level, supported by the measurements reported in Section~\ref{evaluation}.

\section*{Code and data availability}

The framework is TyxonQ 1.4.0 at commit \texttt{e51d61dc8e1eb90c78941fe8c0da4990e2b0489d}, available from \url{https://github.com/QureGenAI-Biotech/TyxonQ}. No real hardware credentials or vendor measurement data are included.
\begingroup
\setlength{\bibsep}{3pt}
\RaggedRight
\bibliographystyle{unsrtnat}
\bibliography{references}

@article{cao2019,
  author = {Cao, Yudong and Romero, Jonathan and Olson, Jonathan P. and Degroote, Matthias and Johnson, Peter D. and Kieferov{\'a}, M{\'a}ria and Kivlichan, Ian D. and Menke, Tim and Peropadre, Borja and Sawaya, Nicolas P. D. and Sim, Sukin and Veis, Libor and Aspuru-Guzik, Al{\'a}n},
  title = {Quantum chemistry in the age of quantum computing},
  journal = {Chemical Reviews},
  volume = {119},
  number = {19},
  pages = {10856--10915},
  year = {2019},
  doi = {10.1021/acs.chemrev.8b00803}
}

@article{mcardle2020,
  author = {McArdle, Sam and Endo, Suguru and Aspuru-Guzik, Al{\'a}n and Benjamin, Simon C. and Yuan, Xiao},
  title = {Quantum computational chemistry},
  journal = {Reviews of Modern Physics},
  volume = {92},
  pages = {015003},
  year = {2020},
  doi = {10.1103/RevModPhys.92.015003}
}

@misc{qiskit2024,
      title={Quantum computing with {Q}iskit}, 
      author={Ali Javadi-Abhari and Matthew Treinish and Kevin Krsulich and Christopher J. Wood and Jake Lishman and Julien Gacon and Simon Martiel and Paul D. Nation and Lev S. Bishop and Andrew W. Cross and Blake R. Johnson and Jay M. Gambetta},
      year={2024},
      eprint={2405.08810},
      archivePrefix={arXiv},
      primaryClass={quant-ph},
      doi={10.48550/arXiv.2405.08810}
}

@misc{cirq,
  title={Cirq}, 
  author={Cirq Developers},
  url={https://zenodo.org/doi/10.5281/zenodo.4062499}, 
  DOI={10.5281/ZENODO.4062499}, 
  abstractNote={Python package for writing, manipulating, and running quantum circuits on quantum computers and simulators.}, 
  publisher={Zenodo},  
  year={2026}
}

@article{pennylane,
  author    = {Ville Bergholm and Josh Izaac and Maria Schuld and Christian Gogolin and Carsten Blank and Keri McKiernan and Nathan Killoran},
  title     = {PennyLane: Automatic differentiation of hybrid quantum-classical computations},
  journal   = {arXiv preprint arXiv:1811.04968},
  year      = {2018},
}

@misc{tensorcircuit,
  author = {{TensorCircuit Team}},
  title = {{TensorCircuit-NG}},
  year = {2026},
  url = {https://github.com/tensorcircuit/tensorcircuit-ng},
  note = {Version 1.10.0}
}

@misc{qiskit-nature,
  author = {{Qiskit Nature Development Team}},
  title = {{Qiskit Nature}},
  year = {2026},
  url = {https://github.com/qiskit/qiskit-nature},
  note = {Version 0.8.0}
}

@misc{pennylane-qchem,
  author = {{PennyLane Team}},
  title = {{PennyLane} quantum chemistry module},
  year = {n.d.},
  url = {https://docs.pennylane.ai/en/stable/introduction/chemistry.html},
}

@misc{tencirchem,
  author = {{TenCirChem Contributors}},
  title = {{TenCirChem}},
  year = {n.d.},
  url = {https://github.com/tencent-quantum-lab/tencirchem},
}

@misc{dononelli2026,
  author = {Dononelli, Wilke},
  title = {Integration of variational quantum algorithms into atomistic simulation workflows},
  howpublished = {arXiv preprint arXiv:2602.02695},
  year = {2026},
}

@misc{nirmal2022,
  author = {Nirmal, M. R. and Sarkar, Shampa and Nambiar, Manoj},
  title = {An interface for variational quantum eigensolver based energy ({VQE-E}) and force ({VQE-F}) calculator to atomic simulation environment ({ASE})},
  howpublished = {arXiv preprint arXiv:2209.13876},
  year = {2022},
}

@article{qfm,
  author = {Shajan, Akhil and Kaliakin, Danil and Liang, Fangchun and Pellegrini, Thaddeus and Doga, Hakan and Bhowmik, Subhamoy and Das, Susanta and Mezzacapo, Antonio and Motta, Mario and Merz, Jr., Kenneth M.},
  title = {Molecular Quantum Computations on a Protein},
  journal = {Journal of Chemical Theory and Computation},
  volume = {22},
  number = {12},
  pages = {6041--6056},
  year = {2026},
  doi = {10.1021/acs.jctc.6c00364},
  note = {{Qiskit Community}. quantum-fragment-methods (software). URL \url{https://github.com/qiskit-community/quantum-fragment-methods}.}
}

@misc{ewfgeom,
  author = {Kaliakin, D. and Shajan, A. and Liang, F. and Li, Z. and Merz, Jr., K. M.},
  title = {Quantum-centric geometry optimization with wave-function-based embedding},
  howpublished = {arXiv preprint arXiv:2607.16410},
  year = {2026},
  url = {https://arxiv.org/abs/2607.16410},
  note = {{MIQuLab}/ewf-geom-opt (software), URL \url{https://github.com/MIQuLab/ewf-geom-opt}.}
}

@misc{ase,
  author = {{ASE Developers}},
  title = {Atomic simulation environment},
  year = {n.d.},
  url = {https://ase-lib.org},
}

@misc{ipi,
  author = {{i-PI Developers}},
  title = {{i-PI}},
  year = {n.d.},
  url = {https://ipi-code.org},
}

@misc{openmm,
  author = {{OpenMM Team}},
  title = {{OpenMM}},
  year = {n.d.},
  url = {https://openmm.org},
}

@misc{lammps,
  author = {{LAMMPS Developers}},
  title = {{LAMMPS}},
  year = {n.d.},
  url = {https://www.lammps.org},
}

@misc{mdi,
  author = {{MolSSI}},
  title = {{MolSSI Driver Interface}},
  year = {n.d.},
  url = {https://molssi-mdi.github.io/MDI_Library/},
}

@misc{pennylane-pulse,
  author = {Kottmann, Korbinian},
  title = {Differentiable pulse programming with qubits in {PennyLane}},
  year = {2023},
  url = {https://pennylane.ai/demos/tutorial_pulse_programming101},
}

@misc{qiskit-pulse-removal,
  author = {{IBM Quantum}},
  title = {{Qiskit SDK 2.0} release notes},
  year = {2025},
  url = {https://quantum.cloud.ibm.com/docs/api/qiskit/release-notes/2.0},
}

@article{sqd2024,
  author = {Robledo-Moreno, Javier and Motta, Mario and Haas, Holger and others},
  title = {Chemistry beyond the scale of exact diagonalization on a quantum-centric supercomputer},
  journal = {Science Advances},
  volume = {11},
  number = {25},
  pages = {eadu9991},
  year = {2025},
  doi = {10.1126/sciadv.adu9991}
}

@misc{vayesta,
  author = {{Booth Group}},
  title = {{Vayesta 1.0.1}},
  year = {2022},
  url = {https://github.com/BoothGroup/Vayesta/tree/v1.0.1}
}

@article{lucj2023,
  author = {Motta, Mario and Sung, Kevin J. and Whaley, K. Birgitta and Head-Gordon, Martin and Shee, James},
  title = {Bridging physical intuition and hardware efficiency for correlated electronic states: the local unitary cluster {Jastrow} ansatz for electronic structure},
  journal = {Chemical Science},
  volume = {14},
  number = {40},
  pages = {11213--11227},
  year = {2023},
  doi = {10.1039/D3SC02516K}
}

@misc{ffsim,
  author = {{ffsim developers}},
  title = {{ffsim}},
  year = {n.d.},
  url = {https://github.com/Qiskit/ffsim},
}

@article{ase2017,
  author = {Larsen, Ask Hjorth and others},
  title = {The atomic simulation environment---a {Python} library for working with atoms},
  journal = {Journal of Physics: Condensed Matter},
  volume = {29},
  pages = {273002},
  year = {2017},
  doi = {10.1088/1361-648X/aa680e}
}

@article{ipi2019,
  author = {Kapil, Venkat and others},
  title = {{i-PI 2.0}: A universal force engine for advanced molecular simulations},
  journal = {Computer Physics Communications},
  volume = {236},
  pages = {214--223},
  year = {2019},
  doi = {10.1016/j.cpc.2018.09.020}
}

@article{openmm2017,
  author = {Eastman, Peter and others},
  title = {{OpenMM 7}: Rapid development of high performance algorithms for molecular dynamics},
  journal = {PLoS Computational Biology},
  volume = {13},
  number = {7},
  pages = {e1005659},
  year = {2017},
  doi = {10.1371/journal.pcbi.1005659}
}

@article{lammps2022,
  author = {Thompson, Aidan P. and others},
  title = {{LAMMPS} -- a flexible simulation tool for particle-based materials modeling at the atomic, meso, and continuum scales},
  journal = {Computer Physics Communications},
  volume = {271},
  pages = {108171},
  year = {2022},
  doi = {10.1016/j.cpc.2021.108171}
}

@article{aiida2020,
  author = {Huber, Sebastiaan P. and Zoupanos, Spyros and Uhrin, Martin and others},
  title = {{AiiDA 1.0}, a scalable computational infrastructure for automated reproducible workflows and data provenance},
  journal = {Scientific Data},
  volume = {7},
  pages = {300},
  year = {2020},
  doi = {10.1038/s41597-020-00638-4}
}

@article{fireworks2015,
  author = {Jain, Anubhav and Ong, Shyue Ping and Chen, Wei and others},
  title = {{FireWorks}: a dynamic workflow system designed for high-throughput applications},
  journal = {Concurrency and Computation: Practice and Experience},
  volume = {27},
  number = {17},
  pages = {5037--5059},
  year = {2015},
  doi = {10.1002/cpe.3505}
}

@article{qmflows2019,
  author = {Zapata, Felipe and Ridder, Lars and Hidding, Johan and Jacob, Christoph R. and Infante, Ivan and Visscher, Lucas},
  title = {{QMflows}: A tool kit for interoperable parallel workflows in quantum chemistry},
  journal = {Journal of Chemical Information and Modeling},
  volume = {59},
  number = {7},
  pages = {3191--3197},
  year = {2019},
  doi = {10.1021/acs.jcim.9b00384}
}

@misc{quacc,
  author = {Rosen, Andrew S.},
  title = {{quacc}: {Quantum Accelerator} (software)},
  year = {n.d.},
  url = {https://github.com/Quantum-Accelerators/quacc},
}

@article{pyscf2020,
  author = {Sun, Qiming and others},
  title = {Recent developments in the {PySCF} program package},
  journal = {Journal of Chemical Physics},
  volume = {153},
  pages = {024109},
  year = {2020},
  doi = {10.1063/5.0006074}
}

@article{openfermion2020,
  author = {McClean, Jarrod R. and others},
  title = {{OpenFermion}: the electronic structure package for quantum computers},
  journal = {Quantum Science and Technology},
  volume = {5},
  pages = {034014},
  year = {2020},
  doi = {10.1088/2058-9565/ab8ebc}
}

@article{zhang2023tensorcircuit,
  title={Tensorcircuit: a quantum software framework for the nisq era},
  author={Zhang, Shi-Xin and Allcock, Jonathan and Wan, Zhou-Quan and Liu, Shuo and Sun, Jiace and Yu, Hao and Yang, Xing-Han and Qiu, Jiezhong and Ye, Zhaofeng and Chen, Yu-Qin and others},
  journal={Quantum},
  volume={7},
  pages={912},
  year={2023},
  publisher={Verein zur F{\"o}rderung des Open Access Publizierens in den Quantenwissenschaften}
}

@article{li2023tencirchem,
  title={TenCirChem: An efficient quantum computational chemistry package for the NISQ era},
  author={Li, Weitang and Allcock, Jonathan and Cheng, Lixue and Zhang, Shi-Xin and Chen, Yu-Qin and Mailoa, Jonathan P and Shuai, Zhigang and Zhang, Shengyu},
  journal={Journal of Chemical Theory and Computation},
  volume={19},
  number={13},
  pages={3966--3981},
  year={2023},
  publisher={ACS Publications}
}
\endgroup

\clearpage
\appendix
\section{Scanner contract}\label{appendix-a.-scanner-contract}

The scanner is the unit of composition. Its contract is intentionally small.

\textbf{Inputs.}

{\def\LTcaptype{none} 
\begin{longtable}[]{@{}
  >{\raggedright\arraybackslash}p{(\linewidth - 6\tabcolsep) * \real{0.2500}}
  >{\raggedright\arraybackslash}p{(\linewidth - 6\tabcolsep) * \real{0.2500}}
  >{\raggedright\arraybackslash}p{(\linewidth - 6\tabcolsep) * \real{0.2500}}
  >{\raggedright\arraybackslash}p{(\linewidth - 6\tabcolsep) * \real{0.2500}}@{}}
\toprule\noalign{}
\begin{minipage}[b]{\linewidth}\raggedright
Argument
\end{minipage} & \begin{minipage}[b]{\linewidth}\raggedright
Type
\end{minipage} & \begin{minipage}[b]{\linewidth}\raggedright
Units / convention
\end{minipage} & \begin{minipage}[b]{\linewidth}\raggedright
Notes
\end{minipage} \\
\midrule\noalign{}
\endhead
\bottomrule\noalign{}
\endlastfoot
\texttt{atom} & \texttt{str} or \texttt{list} & \AA{} by default, or \texttt{unit} & PySCF atom syntax \\
\texttt{basis} & \texttt{str} & n/a & passed to PySCF \\
\texttt{charge} & \texttt{int} & n/a & total charge \\
\texttt{spin} & \texttt{int} & \(N_\alpha-N_\beta\) & PySCF spin convention \\
\texttt{unit} & \texttt{"Angstrom"} or \texttt{"Bohr"} & n/a & geometry unit \\
\texttt{active\_space} & \texttt{(n\_electrons,\ n\_orbitals)} & n/a & spatial orbitals \\
\texttt{method} & \texttt{"sqd"}, \texttt{"uccsd"}, \texttt{"rouccsd"}, \texttt{"hea"} & n/a & default \texttt{"sqd"} \\
\texttt{sampler} & callable & n/a & SQD: sampler or explicit \texttt{ci\_strs} \\
\texttt{solver\_kwargs} & \texttt{dict} & n/a & method options; \texttt{runtime}, \texttt{provider}, \texttt{device}, \texttt{shots} for the device path \\
\texttt{mm\_charges} & \texttt{(coords,\ charges)} & same \texttt{unit} & electrostatic embedding \\
\texttt{mm\_lattice} & diagonal 3$\times$3 array & same \texttt{unit} & orthorhombic Ewald cell \\
\texttt{rcut\_ewald}, \texttt{rcut\_hcore} & float & same \texttt{unit} & Ewald cutoffs \\
\texttt{subspace} & \texttt{"frozen"}, \texttt{"refresh"}, \texttt{"adaptive"} & n/a & CASCI--SQD guard only \\
\texttt{allow\_}\allowbreak\texttt{discontinuous} & bool & n/a & explicit opt-in for non-smooth surfaces \\
\end{longtable}
}

\textbf{Output.} \texttt{scanner(geometry\_bohr)\ -\textgreater{}\ (E,\ dE)} with \texttt{E} in Hartree and \texttt{dE} in Hartree/Bohr, shape \texttt{(n\_atoms,\ 3)}. Bare coordinate arrays use Bohr; a supplied \texttt{Mole} contributes its Bohr coordinates. For periodic embedding, \(r_{\mathrm{cut,Ewald}}<L_{\min}\) and \(r_{\mathrm{QM}}<r_{\mathrm{cut,hcore}}<L_{\min}/2\), in the declared input unit, are enforced by the implementation.

\textbf{Error semantics.} An unsupported method raises at construction. For the CASCI--SQD scanner, \texttt{refresh}/\texttt{adaptive} raises unless \texttt{allow\_discontinuous=True}. Missing optional dependencies may raise an \texttt{ImportError}. The contract does not silently select a different solver method; the documented zero-shot VQE numerical fast path is an execution-mode choice.

\section{Result schema and bit order}\label{appendix-b.-result-schema-and-bit-order}

A device result is normalized to:

\begin{verbatim}
{
  "result": {"00": 512, "11": 488},     # counts
  "result_meta": {...},                 # provider payload
  "uni_status": "completed",            # lifecycle status
  "error": null
}
\end{verbatim}

Postprocessing adds a \texttt{postprocessing} field with the method name and its result. Expectations are computed from counts or, for analytic simulator execution, probabilities; they are not returned by the driver. In the tested \texttt{Circuit} interface, \texttt{device(shots=0).run()} still returns 1024 sampled shots, whereas explicit \texttt{run(shots=0)} returns analytic probabilities. This argument-handling behavior is distinct from the UCCSD scanner's zero-shot numerical fast path; the scanner check does not validate the general execution interface.

\textbf{Bit order.} The state-vector convention places qubit 0 at the most significant bit, and shared counts postprocessing assumes q0-leftmost strings. The LUCJ--SQD example applies explicit reversals within the \(\alpha\)/\(\beta\) blocks when converting samples to determinant occupations. Provider-specific count normalization has not been established for every hardware adapter; preserving a provider's measurement-order metadata is not the same as reordering its counts. An in-process mock checks an already normalized fixture without credentials: counts \texttt{\{"01":\ 1000\}} give \(Z_0=+1\), \(Z_1=-1\), and \(Z_0Z_1=-1\). This tests shared postprocessing, not a native provider payload.

\section{Method and option matrix}\label{appendix-c.-method-and-option-matrix}

\textbf{Table A1. Solver methods and option sets.}

{\def\LTcaptype{none} 
\begin{longtable}[]{@{}
  >{\raggedright\arraybackslash}p{(\linewidth - 8\tabcolsep) * \real{0.2000}}
  >{\raggedright\arraybackslash}p{(\linewidth - 8\tabcolsep) * \real{0.2000}}
  >{\raggedright\arraybackslash}p{(\linewidth - 8\tabcolsep) * \real{0.2000}}
  >{\raggedright\arraybackslash}p{(\linewidth - 8\tabcolsep) * \real{0.2000}}
  >{\raggedright\arraybackslash}p{(\linewidth - 8\tabcolsep) * \real{0.2000}}@{}}
\toprule\noalign{}
\begin{minipage}[b]{\linewidth}\raggedright
Method
\end{minipage} & \begin{minipage}[b]{\linewidth}\raggedright
Reference
\end{minipage} & \begin{minipage}[b]{\linewidth}\raggedright
Runtime options
\end{minipage} & \begin{minipage}[b]{\linewidth}\raggedright
Key options
\end{minipage} & \begin{minipage}[b]{\linewidth}\raggedright
Force / MD boundary
\end{minipage} \\
\midrule\noalign{}
\endhead
\bottomrule\noalign{}
\endlastfoot
\texttt{uccsd} & RHF & numeric/device & \texttt{config\_}\allowbreak\texttt{function}, \texttt{device\_opts} & smoothness and noise checks needed \\
\texttt{rouccsd} & ROHF scanner & numeric/device & spin, \texttt{config\_}\allowbreak\texttt{function}, \texttt{device\_opts} & smoothness and noise checks needed \\
\texttt{hea} & RHF/ROHF scanner & numeric/device & \texttt{n\_layers}, mapping, \texttt{device\_opts} & smoothness and noise checks needed \\
\texttt{sqd} & RHF/ROHF scanner & sampler or fixed strings & subspace, \texttt{ci\_strs}, sampler options & CASCI scanner freezes subspace \\
LUCJ + SQD & RHF example & sampler & \texttt{n\_layers}, shots, sampler & EWF example resamples \\
EWF & RHF/UHF & classical / callback & fragments, bath, CCSD/FCI or callback & closed-shell approximate-gradient scanner \\
PCM--SQD & fixed orbitals & sampler & solvent, outer iterations & energy/density only; no nuclear-gradient scanner \\
\end{longtable}
}

The table distinguishes the CASCI scanner reference from the broader solver implementation. VQE-family options are passed in \texttt{solver\_kwargs}, and sampling options are set on the sampler. At zero shots the VQE device configuration uses the numerical fast path. Convergence alone does not establish MD suitability; method-specific smoothness and force-noise checks remain necessary. The difference follows from the quantum step, which is expectation estimation in one case and sampling in the other. EWF has its own solver and callback configuration; its conventional CCSD/FCI methods are not additional \texttt{qc\_scanner} registry entries. Numerical evaluation of a quantum UCCSD ansatz is distinct from conventional classical CCSD.

\section{Provider status}\label{appendix-d.-provider-status}

\textbf{Table A2. Provider status at the version studied.}

{\def\LTcaptype{none} 
\begin{longtable}[]{@{}lllll@{}}
\toprule\noalign{}
Provider & Type & Driver & Credentials & Artifact \\
\midrule\noalign{}
\endhead
\bottomrule\noalign{}
\endlastfoot
\texttt{simulator::statevector} & local & complete & none & IR \\
\texttt{simulator::density\_matrix} & local & complete & none & IR \\
\texttt{simulator::matrix\_product\_state} & local & complete & none & IR \\
\texttt{tyxonq::homebrew\_s2} & hardware & adapter & token & QASM2 \\
\texttt{qcos} & hardware & adapter & access/secret key & vendor object \\
\texttt{quafu} & hardware & adapter & token & QASM2 \\
\texttt{lqcloud} & hardware & adapter & SDK/token & SDK object \\
\texttt{guodun} & hardware & adapter & SDK & QCIS \\
\texttt{ibm} & hardware & skeleton & n/a & excluded \\
\end{longtable}
}

The IBM skeleton is excluded. The five counted hardware adapters are tyxonq, qcos, quafu, lqcloud, and guodun. ``Adapter'' denotes submission/result source code, not complete lifecycle coverage or tested hardware acceptance; in particular, QCOS cancellation is not implemented.

\section{Artifact examples}\label{appendix-e.-artifact-examples}

The three-qubit export/import check retained from the reference workflow is:

\begin{verbatim}
import tyxonq as tq
from tyxonq.compiler.api import compile as tq_compile
from tyxonq.compiler.pulse_compiler.native.qasm3_importer import qasm3_to_circuit
from tyxonq.compiler.gate_compiler.interop.qiskit_interop import qasm_to_ir

circ = (
    tq.Circuit(3).h(0).cx(0, 1).rz(2, 0.3)
    .measure_z(0).measure_z(1).measure_z(2)
)
qasm2 = tq_compile(circ, output="qasm2")["compiled_source"]
qasm3 = tq_compile(circ, output="qasm3")["compiled_source"]
assert qasm_to_ir(qasm2).ops == circ.ops
assert qasm3_to_circuit(qasm3).ops == circ.ops
\end{verbatim}

The QASM2 export of a two-qubit Bell circuit:

\begin{verbatim}
OPENQASM 2.0;
include "qelib1.inc";
qreg q[2];
creg c[2];
h q[0];
cx q[0],q[1];
measure q[0]->c[0];
measure q[1]->c[1];
\end{verbatim}

The QASM3 export of the same circuit uses \texttt{qubit{[}2{]}\ q;} and individual \texttt{measure} statements. The TQASM 0.2 excerpt for a one-qubit \texttt{H} contains \texttt{defcalgrammar\ "openpulse"}, a frame declaration, and a DRAG calibration; the full artifact is 438 characters in the version studied. The JSON IR is a direct serialization of \texttt{num\_qubits}, \texttt{ops}, \texttt{metadata}, and \texttt{instructions}.

\section{Reproduction}\label{appendix-f.-reproduction}

Run from the extracted supplementary bundle's root, using Python 3.12. The source checkout is separate from the supplementary scripts:

\begin{verbatim}
git clone https://github.com/QureGenAI-Biotech/TyxonQ TyxonQ
git -C TyxonQ checkout e51d61dc8e1eb90c78941fe8c0da4990e2b0489d
python -m venv .venv
. .venv/bin/activate
python -m pip install -e ./TyxonQ
python -m pip install -r paper_code/requirements.txt

# benchmarks, cases A/B/C, manuscript snippets, framework probes
python paper_code/run_reproduction.py --source TyxonQ

# figures from recorded JSON; writes the manuscript PNGs
python analysis/make_figures.py

# optional full core suite 
# one cqlib-related failure, requires provider-specific validation on physical hardware
python paper_code/run_reproduction.py --source TyxonQ --core-tests
\end{verbatim}

The runner sets numerical-library thread counts to one and records the source commit, dependency versions, commands, and return codes. A nonzero exit status must be inspected rather than treated as a completed validation. Existing fourth-version outputs are archived on rerun; timings can change with machine load. Figure generation reads the recorded runtime, artifact, and overhead JSON, rather than embedding timing constants in the plotting code.

Executed versions: TyxonQ 1.4.0, Python 3.12.14, Qiskit 2.6.1, PySCF 2.14.0, NumPy 2.5.4, SciPy 1.18.1, PyTorch 2.15.0, ASE 3.29.0, OpenFermion 1.8.1, and Matplotlib 3.11.2 on macOS/ARM64. These checks used an existing environment matching the pins; a fresh installation was not independently tested. No hardware job was submitted. The core suite returned 763 passed, one failed (optional \texttt{cqlib} unavailable), and 21 skipped.
\end{document}